\documentclass[lettersize,journal]{IEEEtran}
\usepackage{amsmath,amsfonts}
\usepackage{array}
\usepackage{textcomp}
\usepackage{stfloats}
\usepackage{url}
\usepackage{verbatim}
\usepackage{graphicx}
\usepackage{cite}
\usepackage{multirow}
\usepackage{soul} 
\usepackage{colortbl} 
\graphicspath{{TRIM_fig/}}
\DeclareGraphicsExtensions{.eps,.png}
\usepackage{caption}
\usepackage{subcaption}
\usepackage[ruled,vlined]{algorithm2e}
\makeatletter
\newcommand{\removelatexerror}{\let\@latex@error\@gobble}
\makeatother
\usepackage[finalizecache=false,frozencache=true,cachedir= _minted-output]{minted}

\begin{document}

\title{TRIM: A Design Space Exploration Model for Deep Neural Networks Inference and Training Accelerators}

\author{Yangjie~Qi,~\IEEEmembership{Student~Member,~IEEE,}
        Shuo~Zhang,~\IEEEmembership{Student~Member,~IEEE,}
        and~Tarek~M.~Taha,~\IEEEmembership{Member,~IEEE}
\thanks{This paper was produced by the Parallel Cognitive Systems Laboratory at University of Dayton. They are in Dayton, OH.}
\thanks{Manuscript received April xx, xxxx; revised August xx, xxxx.}}

\markboth{Journal of \LaTeX\ Class Files,~Vol.~14, No.~8, August~2021}%
{Shell \MakeLowercase{\textit{et al.}}: A Sample Article Using IEEEtran.cls for IEEE Journals}

\IEEEpubid{0000--0000/00\$00.00~\copyright~2021 IEEE}

\maketitle

\begin{abstract}
There is increasing demand for specialized hardware for training deep neural networks, both in edge/IoT environments and in high performance computing systems. The design space of such hardware is very large due to the wide range of processing architectures, deep neural network configurations, and dataflow options. This makes developing deep neural network processors quite complex, especially for training. We present TRIM, an infrastructure to help hardware architects explore the design space of deep neural network accelerators for both inference and training in the early design stages. The model evaluates at the whole network level, considering both inter-layer and intra-layer activities. Given applications, essential hardware specifications, and a design goal, TRIM can quickly explore different hardware design options, select the optimal dataflow and guide new hardware architecture design. We validated TRIM with FPGA-based implementation of deep neural network accelerators and ASIC-based architectures. We also show how to use TRIM to explore the design space through several case studies. TRIM is a powerful tool to help architects evaluate different hardware choices to develop efficient inference and training architecture design. Experimental results show that TRIM is a powerful tool for rapidly exploring the design space of DNN architectures for training and inference.
\end{abstract}


\section{Introduction}\label{sec:introduction}
\IEEEPARstart{D}{eep} Neural Networks (DNNs)~\cite{Lecun2015a} are being used in a wide variety of application domains, including computer vision, natural language processing, big data analysis, among others. The success of DNNs is also leading to intensive studies on DNN designs in different scenarios, from the data center to IoT. Most of these studies are focused on the DNN inference accelerator. Recent DNN application use cases are showing the need for various types of DNN training accelerators.


For very large deep learning models, which are generally processed in data centers, the training time and energy costs are becoming critical limiting factors. For example, GPT-3, which has 175 billion parameters, would take 355 years and \$4,600,00 for training using one Tesla V100 cloud instance~\cite{Li2020}. Therefore further investigations into more optimized DNN training accelerators for cluster environments are needed. On the other hand, training DNNs on mobile devices is also gaining attention through Federated Learning from Google~\cite{Hard2018}. In this case, the user does not share their private data with the cloud/server, and instead trains a network on their local system. A batch of federated learning applications have already been built and applied to cell phones~\cite{Hard2018}. Therefore, an energy-efficient DNN accelerator chip would be of great help for improving the user's experience. Moreover, training in real-time is becoming increasingly important since many edge devices are collecting new data on their own, such as IoT devices~\cite{Kukreja2019} and robotics~\cite{Su2020}. The input data of these devices are usually sequential and change once in a while. Thus they require the processing hardware to be low-power with online learning capabilities.

Different scenarios require DNN hardware accelerators to offer significantly distinct requirements, including performance, power, latency, chip area, and flexibility. The design space for such DNN accelerators is complicated by the wide range of network architectures, numerous possible dataflows, and various hardware choices. Investigating this large hardware design space is currently an ad hoc and laborious process that requires significant time and expense to evaluate different design options. Many accelerators have been proposed for deep learning. These include systems for inference~\cite{Lu2017,Reagen2016,Han2016,Chen2017b} and on-chip training~\cite{Luo2017a,Schuiki2018a}. They share a similar architecture: a specialized memory hierarchy connected with an array of multiply-accumulate units. The differences between the designs come from their dataflows, which define the method of data partitioning and the order of computations. These accelerators are generally co-designed based on their dataflows to satisfy data cache and transformation bandwidth requirements.

\IEEEpubidadjcol
Several groups have proposed analytical models to help explore the design space of the inference DNN accelerator. Timeloop~\cite{Parashar2019b} analyzes the data movements and memory access patterns to estimate the performance of the DNN inference accelerator. MAESTRO~\cite{Kwon2020} utilizes data-centric directives and their co-designed analytical cost model to explore the design space of DNN inference accelerators. These two models give time and energy estimates based on the individual layers of a DNN. They only consider intra-layer (single layer level) workloads, and not inter-layer(cross-layer level) workloads. This means these models cannot achieve optimal results at the network level. AutoDNNChip~\cite{Xu2020} recognized the importance of cross-level optimization and utilized a graph-based representation to help predict the performance of accelerators. However, their scope is also limited to inference DNN accelerators. The comparisons between TRIM and all the above analytical models are shown in Table~\ref{table:comp}. TRIM is the only analytical model that supports all three phases of training tasks. It is also the only analytic model that is considered both intra-layer workloads and inter-layer workloads.
\IEEEpubidadjcol

\begin{table}[htbp]
\caption{TRIM compare with other analytical models}
\begin{center}
\begin{tabular}{|c|c|c|c|c|c|c|}
\hline
&\multicolumn{4}{c|}{Intra-layer}&\multicolumn{2}{c|}{Inter-layer} \\
&\multicolumn{4}{c|}{Workloads}&\multicolumn{2}{c|}{Workloads} \\
\hline
& \multirow{2}{*}{FW} & \multirow{2}{*}{BW} & WG \& & \multirow{2}{*}{Pooling} & \multirow{2}{*}{DP$^1$} & \multirow{2}{*}{DD$^2$} \\
& & &  Update & & & \\
\hline
TRIM & Y & Y & Y & Y & Y & Y \\
\hline
Timeloop & Y & N & N & N & N & N \\
\hline
MAESTRO & Y & N & N & Y & N & N \\
\hline
AutoDNNChip & Y & N & N & Y & N & N \\
\hline
\end{tabular}
\label{table:comp}
\end{center}
\footnotesize{$^1$ DP: Data Preprocessing $^2$ DD: Data Dependency (Activation Caching)}\\
\end{table}

Our review of the literature shows a lack of analytical models and design space exploration tools designed for DNN training accelerators. On the one hand, compared with DNN inference accelerators, training accelerators should be able to process more complicated intra-layer patterns. For example, the kernel height/width of the inference phase is mainly 3, 5, 7, and 11. In contrast, in the weight gradient phase, kernel height/width could be up to 220 (for AlexNet layer 1). On the other hand, intra-layer optimizations play more critical roles in the training phase, and the workloads are more complex than inference inter-layer workloads. Those training workloads required extra memory and significant data movement energy than inference tasks, which must be considered in the model to make sure the architectures have the capability to process training tasks and get network level optimal results (section 3.3).

In this paper, we present TRIM (TRaining archItecture Model for deep networks), an infrastructure to help hardware architects explore the design space of DNN accelerators for training and inference. It considers both intra-layer workloads and inter-layer workloads of DNNs. Given application and hardware specifications, TRIM quickly examines all possible dataflows and estimates time, energy, and area. TRIM also can be used to compare different hardware design options, optimize existing architectures, and guide new hardware architecture designs. The key contributions of this paper can be summarized as follows:

1) TRIM provides an analytical model to estimate the performance and energy of various DNN hardware architectures. TRIM utilizes a very flexible hardware template, which can model a wide range of architectures. TRIM explores the design space of data partition and reuse strategies for each hardware architecture and estimates the optimal performance and energy. This exploration guarantees fair comparisons between different architectures.

2) TRIM supports both inference and training of DNN accelerators. To the best of our knowledge, TRIM is the first infrastructure that can model and explore the design space of DNN architectures for both training and inference. Furthermore, to accurately model training architectures, TRIM explores the design space of the DNN training accelerators at the network level, considering both intra-layer and inter-layer activities, and finds the optimal design choices.

3) We demonstrate how to utilize TRIM to explore the design space of FPGA-based architectures and ASIC-based architectures through two case studies. The results show the pros and cons of different hardware choices and lead to efficient training architecture designs.

\section{TRIM Overview}\label{section:overview}
\begin{figure}[!t]
\centering
\includegraphics[width=3.0in]{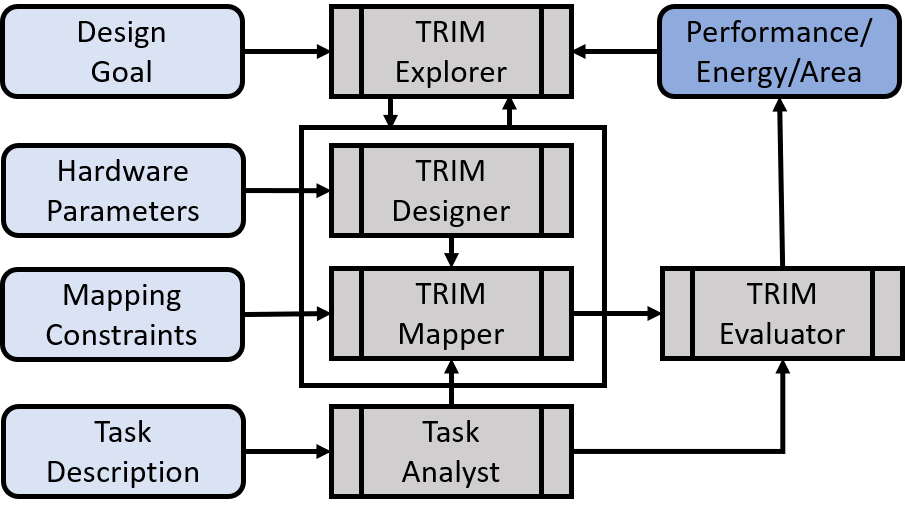}
\caption{Overall view of TRIM model}
\label{fig:model_overview}
\end{figure}

TRIM primarily provides a systematic framework to predict the time, energy, and area for various hardware architectures running different DNN applications for both inference and training. As shown in Fig.~\ref{fig:model_overview}, TRIM takes four inputs: 1) task description, which consists of the DNN network model and corresponding parameters, such as batch size; 2) hardware parameters, which define the system hierarchy and specifications of each hardware component; 3) mapping constraints, which constrain data partitioning and order of computation in the system; 4) design goal, such as the fastest throughput or the lowest energy consumption.

TRIM consists of five components: Task analyst, TRIM designer, TRIM mapper, TRIM evaluator, and TRIM explorer. The task analyst utilizes the task description to generate intra-layer workloads and inter-layer workloads. The TRIM designer generates various hardware descriptions based on the hardware parameters. Each hardware description presents a specific hardware architecture. All hardware descriptions together configure the hardware architecture space. For a given hardware description, the TRIM mapper generates a mapspace for each intra-layer workload. A mapspace consists of multiple mappings, and each mapping defines a specific way in which data is partitioned, staged, and computed across the hardware architecture. The TRIM evaluator estimates the performance and energy of each mapping in the mapspace. Based on the estimates, the TRIM explorer finds the optimal mapping in each mapspace based on the design goal. By combining the optimal mappings and inter-layer workloads, the TRIM evaluator can estimate the performance and energy of each architecture in the architecture space. In the end, the TRIM explorer selects the optimal architecture based on the design goal.

In the remainder of the paper, we present the task analyst and its inputs and outputs in section~\ref{section:task_analyst}. Section~\ref{section:designer} introduces the TRIM designer. The mapping, mapspace, and our method to prune the mapspace are described in section~\ref{section:mapper}. The TRIM evaluator and explorer are described in section~\ref{section:evaluator}. We validate TRIM with FPGA designs and use TRIM to design an FPGA-based DNN accelerator in section~\ref{section:fpga}. Section~\ref{section:asic} is case studies using TRIM to explore ASIC based DNN accelerators design space, while section~\ref{section:conclusion} concludes the paper.

\section{Task Analyst}\label{section:task_analyst}
The task analyst goes through the task description and generates workloads. There are two types of workloads: intra-layer workloads and inter-layer workloads. The intra-layer workloads describe each layer's primary computation operations, such as the 2D convolution in the convolutional (CONV) layer, the 2D pooling in the pooling (POOL) layer, and the matrix-to-matrix multiplication in the fully-connected (FC) layer. The inherent parallelism of the intra-layer workloads makes it possible to achieve performance and energy efficiency in parallel computing architectures. The inter-layer workloads consist of data preprocessing and intermediate activation caching. They do not have many parallel computing opportunities but significantly impact the tasks' performance and energy consumption.

\subsection{Task Description}
\begin{figure}[!t]
\centering
\begin{minted}
[
frame=lines,
framesep=2mm,
baselinestretch=1.2,
fontsize=\scriptsize,
xleftmargin=15pt,
linenos
]
{python}
network_parameters = {
    processing_type = 'Training'
    input_shape = (224,224,3)
    output_shape = 1000
    batch_size = 64
    }

network_model = {
    x = conv2d(in_shape=input_shape,
              out_channel=64,kernel_size=(11,11),
              padding=(2,2),stride=(4,4),
              activation='ReLU'))
    x = pool2d(in_shape=x.shape,
              kernel_size=(3,3), stride=(2,2))
    x = fc(in_shape= x.shape,
          out_channel=output_shape,
          activation='Sigmoid')}
\end{minted}
\caption{An example of the task description}
\label{fig:ex_task_description}
\end{figure}

The task description used in our paper is a simplified TensorFlow-like description, which consists of the network parameters and network model, so it should be easy for TensorFlow users to develop their TRIM model from their code. The network parameters comprise the input shape, output shape, batch size, and processing type (inference or training). The network model is described as multiple layers connected in a specific order, where each layer is defined by a layer type with its corresponding parameters. The CONV layer is defined by in shape, out channels, kernel size, padding size, stride size, and activation type. The FC layer is described by in shape, out channels, and activation types. The parameters of the POOL layer are in shape, kernel size, and stride size. Fig.~\ref{fig:ex_task_description} shows the task description of a three layer network as an example.

The task analyst goes through the task description and generates workloads. In the case of inference, the task analyst generates one intra-layer workload for each layer. For training, however, the number of workloads generated depends on the layer type. For the CONV layer and FC layer, three workloads are generated, which correspond to the forward propagation (FW), backpropagation (BW), and weight gradient and update (WG) phases. The only exception is the first layer of the network, which does not have the BW phase. For the POOL layer, only two workloads are generated, as there is no WG phase needed. For AlexNet~\cite{c}, which has five CONV layers, three fully connected layers, and three POOL layers, the task analyst generates $5+3+3=11$ intra-layer workloads for the inference task, and $(5+3)\times3+3\times2-1=29$ intra-layer workloads for the training task.

\subsection{Intra-layer Workloads}
The intra-layer workload is presented as a nested loop. As shown in Fig.~\ref{nest_loop}, the operation in the CONV layer is described by the height and width of filters (R, S), the height and width of outputs (E, F), input channel size (C), output channel size (M), and batch size (N). These seven parameters are used to construct a nested loop, as shown in Fig.~\ref{nest_loop}. As the multiply–accumulate (MAC) operations occur only in the innermost loop (see Fig.~\ref{nest_loop}), the loops can be nested in any order. The seven parameters, along with the stride sizes (U, V), define the computations and the shapes of the inputs, filters, and outputs. We can also represent fully connected layers, and recurrent layers in the same format as their main computations are matrix-matrix multiplications and matrix-vector multiplications. Matrix-matrix multiplications can be defined similarly by setting R, S, E, and F equal to 1. Matrix-vector multiplications can be represented by placing R, S, E, F, and N equal to 1. The POOL layer is supported and evaluated in the experiments. We consider it an intra-layer workload since it can also be described in a similar nested loop as the CONV layer. For the normalization layer, we assume it would be processed by the CPU or a separate coprocessor, which means it can be modeled as a constant delay in TRIM. Some special connections, such as the residual link, are split into an intra-layer workload and an inter-layer workload, which is considered and evaluated in the ResNet-IM experiments shown in Figs.~\ref{fig:e_break_across_hw} to~\ref{fig:overall_e_break}.


\begin{figure}[!t]
\centering
\begin{minted}
[
frame=lines,
framesep=2mm,
baselinestretch=1.2,
fontsize=\scriptsize,
xleftmargin=15pt,
linenos
]
{python}
for n = 0:N # batch size
 for m = 0:M # out channel
  for c = 0:C # in channel
   for r = 0:R # filter height
    for s = 0:S # filter width
     for e = 0:E # out height
      for f = 0:F # out width
       p = e * u + r # u and v are strides
       q = f * v + s
       output[n,e,f,m] += 
        input[n,p,q,c] * filter[r,s,c,m]
\end{minted}
\caption{Intra-layer workload of TRIM using loop nest format}
\label{nest_loop}
\end{figure}

\subsection{Inter-layer Workloads}
There are two types of inter-layer workloads: data preprocessing and intermediate activations caching.

\textbf{Data preprocessing} is executed before each intra-layer workload. For the inference tasks of DNN, as shown in Eq.\ref{eq_fw}, only padding operations are needed and are usually ignored by inference exploration tools, such as\cite{Parashar2019b}. However, those operations should be taken into consideration, as they consume both time and energy. Especially for the training task, the data preprocessing is executed in each iteration. Eq.~\ref{eq_fw}, Eq.~\ref{eq_wg}, and Eq.~\ref{eq_bw} show the different data preprocessing operations for FW, BW, and WG of the CONV layer.

Furthermore, those preprocessing operations, such as padding and upsampling, would generate zeros. Those zeros are predictable and can be used to estimate the time and energy consumption of those architectures with an early zero detect mechanism. The data preprocessing workloads are presented as operation type, input, and output shape.

\begin{equation}\label{eq_fw}
Y=padding(X)*W
\end{equation}
\begin{equation}\label{eq_wg}
dW=padding(X)*upsampling(dY)
\end{equation}
\begin{equation}\label{eq_bw}
dX=padding(upsampling(dY)*rot180(W^T))
\end{equation}

\begin{figure}[!t]
\centering
\includegraphics[width=3in]{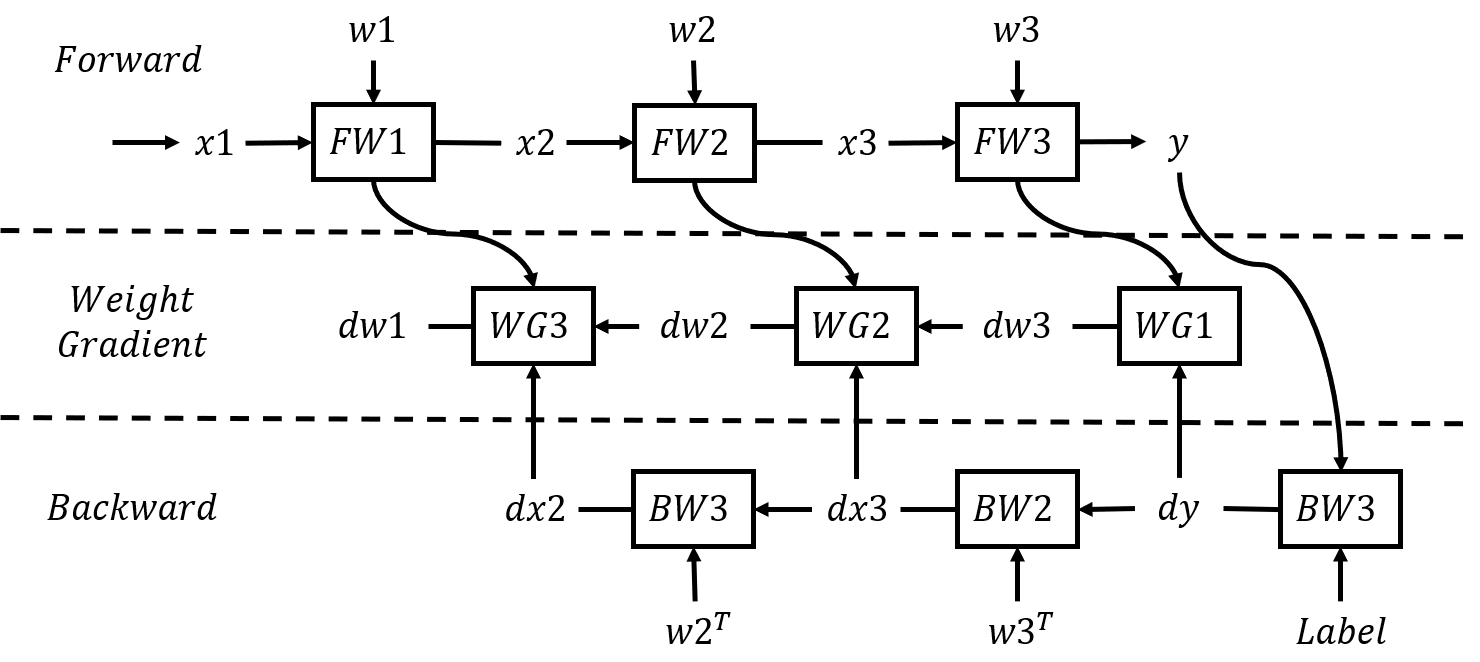}
\caption{Inter-layer data dependency of DNN}
\label{fig:inter_layer}
\end{figure}

\textbf{Intermediate activations caching} is only considered in the training task. As shown in Fig.~\ref{fig:inter_layer}, after we compute the first forward layer (FW1) of the network, the activations $x1$ need to be cached and used as the inputs of WG3 later. Similarly, the activations $x2$ and $x3$ need to be cached and used as the inputs of WG2 and WG1, respectively. After all the FW computations are processed, we compute the first backward phase (BW1) and get the gradient errors ($dy$). The size of those intermediate activations and the timestamp they created and deleted configure the intermediate activations workloads. Data caching consumes both memory resources and energy.  The TRIM mapper considers the memory size used by intermediate activations when validating the mapping, introduced in section~\ref{section:mapper}. The TRIM evaluator computes intermediate activations' energy consumption and adds it to the architecture's overall energy consumption.

\begin{table}[!t]
\renewcommand{\arraystretch}{1.3}
\caption{Architecture Parameters of TRIM}
\label{table:arch_para}
\centering
\begin{tabular}{|c|c|}
\hline
Type        & Architecture Parameters                           \\ \hline
System      & \# of levels; data precision                      \\ \hline
Computation & \# of PEs                                         \\ \hline
Memory      & type; size; usage (inputs/filters/outputs/share)  \\ \hline
Routing     & topology; routing size                            \\ \hline
\end{tabular}
\end{table}

\section{Designer}\label{section:designer}
TRIM Designer utilized the architecture parameter to generate multiple hardware descriptions. Each hardware description presented a specific hardware organization. We utilized a very flexible hardware architecture template, which can be viewed as a tree graph. The main memory is the root, the global buffer (Gbuffer) is an intermediate node, the network-on-chip (NoC) are branches, and the register files (RFs) and processing engines (PEs) are the leaves.  Between the main memory and global buffer, it may have multiple memory or routing levels. It is also possible to have multiple buffers in parallel as different intermediate nodes.

Table~\ref{table:arch_para} shows the architecture parameters we can vary during the design space exploration. Each parameter can be set as multiple values. The designer generates various hardware descriptions by combining different values of each parameter. Table~\ref{table_eyeriss_hw} shows one hardware description generated by the designer to model Eyeriss\cite{Chen2017b}, a popular inference accelerator. As defined in number of levels, it has five hardware levels, which are named processing engine (PE), scratchpads (SP), network-on-chip (NoC), global buffer (Gbuffer), and off-chip memory. The first level is the computation level, which defines the total number of  PEs in the system. The second, fourth, and fifth are memory level. They are different in the memory type and memory size. The value of their usage is shared, which means inputs, filters, and outputs share this memory. It is also possible to specify separate memory used by inputs, filters, and outputs. In that case, at the same hardware level, three separate memories need to be defined. The third level is the routing level, which defines the routing network topology and size.

Using the different values of parameters, the designer generates various hardware architectures. However, we cannot compare those architectures as their performance and energy consumption depend not only on their hardware architectures but also on the mappings. A mapping describes how a workload is executed in a hardware architecture. For a given hardware architecture, millions of possible mappings exist. Therefore, to compare different hardware architectures' performance and energy consumption, we need to ensure that the architectures are evaluated with their optimal mapping.

\begin{table}[!t]
\renewcommand{\arraystretch}{1.3}
\caption{An Example of Eyeriss Hardware Description}
\label{table_eyeriss_hw}
\centering
\begin{tabular}{|c|c|c|c|c|}
\hline
Level              & Type                     & Name                      & Parameter      & Value      \\ \hline
\multirow{2}{*}{}  & \multirow{2}{*}{System}  & \multirow{2}{*}{Arch-1}   & \# of levels   & 5          \\ \cline{4-5}
                   &                          &                           & data precision & fixed 16   \\ \hline
1                  & Computation              & PE                        & \# of PEs      & 256        \\ \hline
\multirow{3}{*}{2} & \multirow{3}{*}{Memory}  & \multirow{3}{*}{SP}       & memory type    & scratchpad \\ \cline{4-5}
                   &                          &                           & memory size    & 520 bytes  \\ \cline{4-5}
                   &                          &                           & usage          & shared     \\ \hline
\multirow{2}{*}{3} & \multirow{2}{*}{Routing} & \multirow{2}{*}{NoC}      & topology       & 2-Level Bus    \\ \cline{4-5}
                   &                          &                           & routing size   & $16\times16$    \\ \hline
\multirow{3}{*}{4} & \multirow{3}{*}{Memory}  & \multirow{3}{*}{Gbuf}     & memory type    & SRAM       \\ \cline{4-5}
                   &                          &                           & memory size    & 108 K      \\ \cline{4-5}
                   &                          &                           & usage          & shared     \\ \hline
\multirow{3}{*}{5} & \multirow{3}{*}{Memory}  & \multirow{3}{*}{Off-chip} & memory type    & DRAM       \\ \cline{4-5}
                   &                          &                           & memory size    & N/A        \\ \cline{4-5}
                   &                          &                           & usage          & shared     \\ \hline
\end{tabular}
\end{table}

\section{Mapper}\label{section:mapper}
\begin{figure}[!t]
\centering
\includegraphics[width=2.8in]{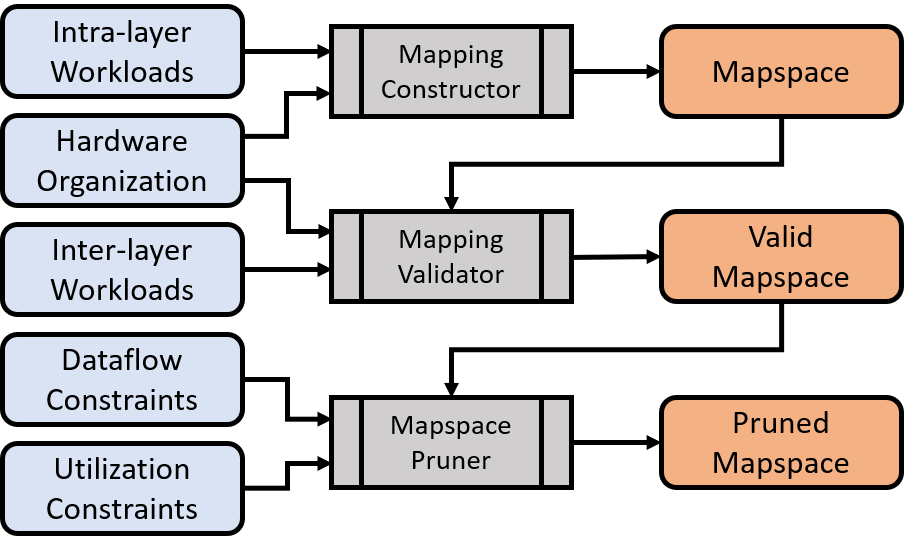}
\caption{Overview of TRIM mapper}
\label{fig:overview_mapper}
\end{figure}

The mapper consists of a mapping constructor, a mapping validator, and a mapspace pruner, as shown in Fig.~\ref{fig:overview_mapper}. The mapping constructor takes intra-layer workloads and hardware organization to create possible mappings. Each mapping utilizes a nested loop format, which describes how data is moved, staged, and computed in given hardware. All possible mappings create the mapspace. The mapping validator computes the size of each hardware component used by the mapping, adjusts with the size used by intermediate activation caching workloads, and compares it with the size constraints listed in the hardware organization.  The mapping that satisfies the size constraints of the hardware organization is called a valid mapping. All possible valid mappings together are the valid mapspace. The number of mappings in the valid mapspace varies from zero to several million. In the case of a mapspace having millions of mappings, exploring the mapspace exhaustively is too time-expensive. Dataflow constraints and utilization constraints are used to prune the valid mapspace. Finally, the pruned mapspace is carried out to the TRIM evaluator and explored by the TRIM explorer.

\subsection{Mapping}
\begin{figure*} 
\centering
\begin{tabular}[t]{cc}
        \begin{tabular}{c}
        \begin{subfigure}[t]{0.4\textwidth}
        \begin{minted}
        [
        frame=lines,
        % framesep=2mm,
        % baselinestretch=1.2,
        fontsize=\scriptsize,
        % xleftmargin=15pt,
        linenos
        ]
        {python}
# Workload 
C = 16
M = 32
for c = 1:C
 for m = 1:M # C and M are loop bounds
  outputs[m] += inputs[c] * filters[c,m]
        \end{minted}
        \caption{}
        \label{fig:ex_mapping_a}
        \end{subfigure} 
        \\
        \begin{subfigure}[t]{0.4\textwidth}
        \begin{minted}
        [
        frame=lines,
        % framesep=2mm,
        % baselinestretch=1.2,
        fontsize=\scriptsize,
        % xleftmargin=15pt,
        linenos
        ]
        {python}
# One possible Mapping
# Off-chip Level
for c4 = 1:1
 for m4 = 1:4
  # GBuf Level
  for m3 = 1:2
   for c3 = 1:2
    # NoC Level
    parallel for c2 = 1:1 # @ x-axis
    parallel for m2 = 1:4 # @ y-axis
     # Spad Level
     for m1 = 1:1
      for c1 = 1:8
       # PE Level
       c = c1+(c2-1)*2+(c3-1)*2+(c4-1)*2
       m = m1+(m2-1)*2+(m3-1)*2+(m4-1)*2
       outputs[m] += filter[c,m] * inputs[c]
        \end{minted}
        \caption{}
        \label{fig:ex_mapping_b}
        \end{subfigure} 
        \\
        \begin{subfigure}[t]{0.4\textwidth}
        \begin{minted}
        [
        frame=lines,
        % framesep=2mm,
        % baselinestretch=1.2,
        fontsize=\scriptsize,
        % xleftmargin=15pt,
        linenos
        ]
        {python}
# One possible Mapping
# Off-chip Level
for c4 = 1:2
 for m4 = 1:2
  # GBuf Level
  for c3 = 1:2
   for m3 = 1:2
    # NoC Level
    parallel for c2 = 1:2 # @ x-axis
    parallel for m2 = 1:2 # @ y-axis
     # Spad Level
     for c1 = 1:2
      for m1 = 1:4
       # PE Level
       c = c1+(c2-1)*2+(c3-1)*2+(c4-1)*2
       m = m1+(m2-1)*2+(m3-1)*2+(m4-1)*2
       outputs[m] += filter[c,m] * inputs[c]
        \end{minted}
        \caption{}
        \label{fig:ex_mapping_c}
        \end{subfigure}
        \end{tabular}
        &
        \begin{tabular}{c}
\begin{subfigure}[t]{0.4\textwidth}
\centering
\includegraphics[height=0.9\textwidth]{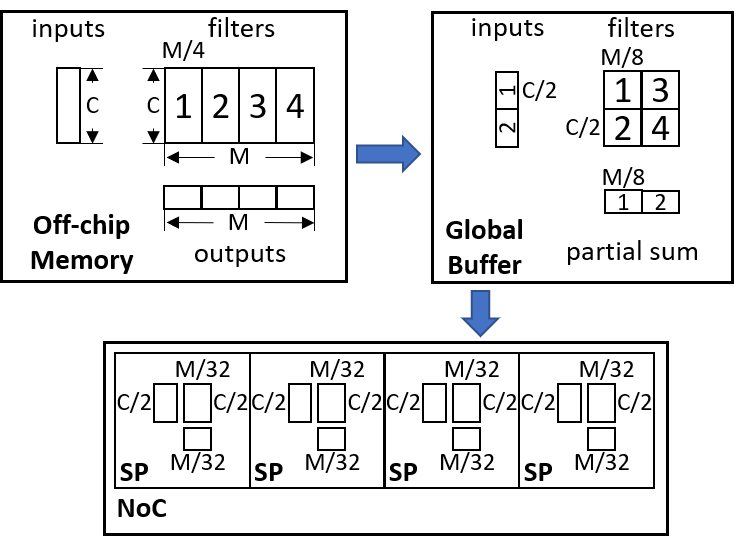}
\caption{}
\label{fig:ex_mapping_d}
\end{subfigure} 
\\
\begin{subfigure}[t]{0.4\textwidth}
\centering
\includegraphics[height=0.9\textwidth]{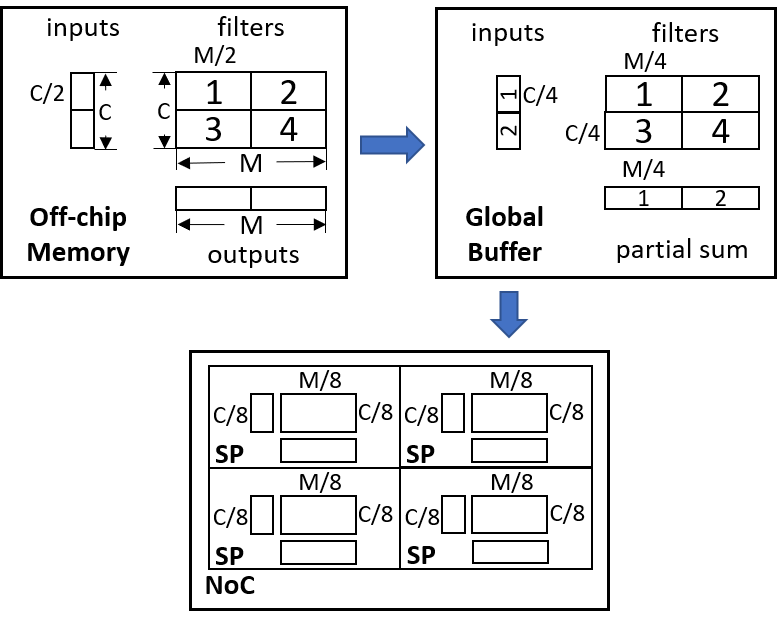}
\caption{}
\label{fig:ex_mapping_e}
\end{subfigure}
        \end{tabular}
\end{tabular}
\caption{An example workload and two possible mappings. (a) is the example workload, (b) and (c) are two example mappings which project the workload onto the hardware defined in Table~\ref{table_eyeriss_hw}. (d) and (e) shows the dataflow based on the mapping in (b) and (c) respectively.}
\label{fig:ex_mapping} 
\end{figure*}

A mapping describes how the data is partitioned, moved, staged, and computed for a workload across a system hierarchy. More specifically, it shows how an intra-layer workload projects onto the hardware organization.  To explain this concept, we will refer to the example in Fig.~\ref{fig:ex_mapping} which shows two possible mappings for a specific workload. Fig.~\ref{fig:ex_mapping_a} shows a vector-matrix multiplication workload, which is a particular case of the workload shown in Fig.~\ref{nest_loop} with M set as 32, C as 16, and all other parameters(N, R, S, E, and F) as 1. Fig.~\ref{fig:ex_mapping_b} and Fig.~\ref{fig:ex_mapping_c} present two mappings using the loop nest format for the workload in Fig~\ref{fig:ex_mapping_a}. Fig.~\ref{fig:ex_mapping_d} and Fig.~\ref{fig:ex_mapping_e} show how the two mappings are visualized in the architectures described in Table~\ref{table_eyeriss_hw}.

Both mappings split the workload into five sub-mappings that correspond to the hardware levels shown in Table~\ref{table_eyeriss_hw}. There are three types of sub-mappings: temporal, spatial, and computational. The temporal level is defined by the \textit{for} loop and used for memory, while the \textit{parallel for} loop defines the spatial level and is used for routing networks. The innermost level is the computational level, which is MAC operations. Temporal and spatial sub-mappings have loops belonging to the same dimensions in workload. As shown in Fig.~\ref{fig:ex_mapping_b} and Fig.~\ref{fig:ex_mapping_c}, the loops c4, c3, c2, c1 correspond to the loop c in workload (Fig.~\ref{fig:ex_mapping_a}), and the product of the loop bounds of loops c1, c2, c3, and c4  is equal to the loop bound of the loop c in the workload.

Each sub-mapping has specific loop bounds and order of loops. The loop bounds in a sub-mapping constrain how the inputs, filters, and outputs are partitioned at the current hardware level. Using the off-chip memory level of two mappings (Fig.~\ref{fig:ex_mapping_b} and Fig.~\ref{fig:ex_mapping_c} line 2 to 4) as example, the inputs and filters are stored in the off-chip memory in the beginning. Assume the global buffer's size is limited, and we cannot load all the data from off-chip memory to the global buffer at once. Thus we should partition the data and deliver them in order. In Fig.~\ref{fig:ex_mapping_d}, the inputs are not partitioned, while the filters are partitioned into four blocks through M dimensions. The partition method is decided by the loop bounds shown in Fig.~\ref{fig:ex_mapping_b} line 3 and 4, where the loop bound is 4 for the M dimension loop and 1 for the C dimension loop. In Fig.~\ref{fig:ex_mapping_e}, the inputs are partitioned into two blocks, and the filters are partitioned into four blocks, but with different shapes. The difference comes from the loop bounds shown in Fig.~\ref{fig:ex_mapping_c} lines 3 and 4, where the loop bounds are 2 for both the M and C dimension loops.

The order of loop dimensions decides the order of data transformation and whether the outputs, inputs, and weights need to be loaded back and sent to the inner hardware level. As shown in Fig.~\ref{fig:ex_mapping_d} and Fig.~\ref{fig:ex_mapping_e}, the data are partitioned using the same method in the global buffer level. The reason for that is the loop bounds of their Gbuf level are the same. However, as the order of their loop dimensions is different, as shown in Fig.~\ref{fig:ex_mapping_b} and Fig.~\ref{fig:ex_mapping_c} lines 6 and 7, the data movement and staging are different. In the first iteration, inputs block-1 and filters block-1 are sent to the NoC level for both mappings. However, for the mapping in Fig.~\ref{fig:ex_mapping_b}, the partial sum results do not need to be loaded back, as the partial sum results of the next iteration belong to the same data block. In contrast, for the mapping in Fig.~\ref{fig:ex_mapping_c}, the partial sum results should be loaded back at the end of the first iteration, as a different partial sum block is computed in the next iteration. In the second iteration, for mapping in Fig.~\ref{fig:ex_mapping_b}, input block-2 and filter block-2 (Fig.~\ref{fig:ex_mapping_d}) are sent to the NoC level. In contrast, for the mapping in Fig.~\ref{fig:ex_mapping_c}, only filter block-2 (Fig.~\ref{fig:ex_mapping_e}) should be delivered as the inputs used for this iteration are the same as the previous one. The order of loop dimensions only affects the temporal sub-mapping. For spatial sub-mappings, the order of loop dimensions is not important, as the data are concurrently delivered to parallel components.

\subsection{Mapping Constructor and Pruner}
The mapping constructor generates all possible mappings for each intra-layer workload by factoring the value of each loop bound of an intra-layer workload across the system levels. The advanced user can also provide custom factor lists, which forces TRIM to follow their factor method to construct mappings. The mapping constructor lists all possible seven loop orders of the loop inside each level. Further, inputs, weights, or outputs may bypass some levels without much reuse. For a given hardware description, the mapping constructor generates a fixed number of mappings for each intra-layer workload. For a specific intra-layer workload, the number of possible mappings is determined by the hardware description. Generally speaking, the mapping constructor generates more mappings for a system with more hierarchy levels, more parallel PEs, or larger memory size.


After a mapping is generated, it is used to compute the different hardware components' required sizes. The required memory size of inter-layer workloads would also be computed and combined with the mapping's memory size needed. The results are validated with the size constraints listed in the hardware organization. If the hardware resource utilization, such as number of PEs and memory utilization, needed by the mapping is less than the amount provided by the hardware organization, we consider the mapping as valid. All invalid mappings would be discarded, and all valid mappings together are the valid mapping space. The number of valid mappings for each layer can vary from zero to several million. A mapspace of this size could potentially be explored but requires significant time to search. We offer two optional methods to prune the valid mapspace and improve the speed of exploration. They are the dataflow constraint method and the utilization constraint method.

Dataflow constraint methods are primarily designed for modeling architectures that have a co-designed dataflow. Most architects develop a dataflow strategy for their architecture to achieve efficiency -- for example, Eyeriss was designed with a row-stationary dataflow in mind. By default, TRIM explores the entire design space with no constraints on dataflow (including the commonly used ones such as row stationary and weight stationary, along with any unique ones). To make TRIM look at a certain dataflow, we could pin some of the for loops in Fig.~\ref{nest_loop} to specific positions. The dataflow constraints (user inputs are shown in Fig.~\ref{fig:model_overview}) are used to select the valid mappings, which comply with specific dataflows rules. At the same time, it can also be used to prune a valid mapspace. However, with more dataflow constraints added, it may exclude optimal mappings in the early stages and lead to inefficient design. We do not add any dataflow constraints to ensure TRIM can find the optimal solution in our architecture design space explorations.

Utilization constraint methods are developed based on our exploration experience using TRIM. We found that the mappings with a high PE utilization are generally faster, in which the PE utilization means the ratio of active PEs divided by the total number of PEs. The mappings with a high utilization rate of memory that is close to PEs are usually more energy-efficient. Thus, we prune the mapspace by setting up the utilization constraints.  For exploration where the design goal is high throughput, we set up the utilization constraints of PE level as 0.75. This means that mappings whose utilization rate of PEs is less than 0.75 would be discarded from the mapspace. When searching energy efficient mappings, we set up the utilization constraints of scratchpads/registers as 0.5. Thus, the mappings whose utilization rate of scratchpads/registers are less than 0.5 would be removed. These two utilization constraint numbers are selected based on our exploration experience. We explore architecture design spaces with constraints and without constraints and got the same optimal mappings.

\section{Evaluator and Explorer}\label{section:evaluator}
TRIM evaluator utilizes mappings and inter-layer workloads as inputs to estimate the performance, energy, and area, as shown in Fig.~\ref{fig:model_overview}. It consists of an activity analyst, a performance model, an energy model, and an area model. The last TRIM component is the TRIM explorer, which tunes all TRIM components to explore the design space and select the optimal architecture.

\subsection{Activity Analyst}
The activity analyst generates the number of MACs, memory accesses, and NoC activities for mappings. The number of MACs is computed as the product of all loop bounds in a mapping. The memory accesses can be counted by using a simulator. However, a simulator would be extremely slow and cannot be used in large design space explorations. As the data transfer and computation in the workload executed are deterministic, we can use a mathematical method to compute the memory accesses quickly and accurately.

The memory accesses are computed using the loop dimensions and bounds belonging to the current hardware level. They are examined in order from inner to outer.  Suppose the current loop examined has dimensions ‘R,’ ‘S,’ ‘M,’ ‘C’ for filters and ‘N,’ ‘M,’ ‘E,’ ‘F’ for outputs. In that case, the memory accesses can be computed as the product of the current loop bound, all unvisited loop bounds, and corresponding filter or output data package size. The memory accesses of inputs are more complex, as the conjunctive packages may overlap. In this case, only part of the data package needs to be transferred. We need to compute the overlap size of two conjunctive iterations in each loop first, then calculate the input memory access. 

The NoC activities are generated based on the \textit{parallel for} loops dimensions and bounds, which define how the inputs, filters, outputs data are transferred in the NoC. If the \textit{parallel for} loop dimension corresponds to batch size (N), output height (E), and output width (F), each node in the NoC gets different input data and the same filter data. The computation results don’t need to be accumulated.  In this case, the filter data is counted as multicast activities, while inputs and outputs are counted as normal data transfer activities. If the \textit{parallel for} loop dimension corresponds to filter height (R), filter width (S), and in channel (C), each node has different input data and filter data, while the computation results belong to the same outputs and need to be accumulated. In this case, the inputs and filters are counted as normal data transfer activities, while the outputs are counted as data transfer with accumulation activities. If the \textit{parallel for} loop dimension corresponds to out channel (M), each node gets the same input data and different filter data,  and the computation results don’t need to be accumulated. Thus the input data and output data are counted as normal data activities, and the filters data are counted as multicast activities. 

\subsection{Performance, Energy, and Area Estimations}
The execution cycle of each intra-layer workload (mapping) is estimated by computing the execution cycles needed for each hardware level first. For PE level, the execution cycle is calculated as the number of MACs divided by the number of PEs and the number pipeline stages of the PEs. For memory and NoC level, the execution cycle is computed as the data transfer size divided by the interface bandwidth.  We assume all hardware levels are operated as a pipeline. Thus, an intra-layer workload's overall time is the maximum of the hardware level's execution cycles. The execution cycle of data preprocessing workload is estimated as the size of output data divided by the memory bandwidth. There is no extra time needed for intermediate activation caching workloads, as the time can be overlapped with the intra-layer workload. The overall performance is computed as the sum of the execution cycle of all intra-layer workloads and data preprocessing workloads.

\begin{figure}[!t]
\removelatexerror
\begin{algorithm}[H]
\SetAlgoLined
\SetKwInOut{Input}{input}\SetKwInOut{Output}{output}

\caption{Design Space Exploration using TRIM}
\Input{design goal, task description, hardware parameters, mapping constraints}
\Output{optimal architecture}\
 intra-layer workloads, inter-layer workloads $\leftarrow$ TaskAnalyst(task description)\;
 Architecture Space $\leftarrow$ Designer(hardware parameters)\;
  \For{each hardware description $\in$  Architecture Space}{
  \For{each intra-layer workload}{
  MapSpace $\leftarrow$ Mapper(hardware organization, intra-layer workloads, mapping constraints)\;
  \For{each mapping $\in$  MapSpace}{
    Performance, Energy,Area $\leftarrow$ Evaluator(mapping)\;
    Evaluate design goal\;
    Update optimalMappings[intra-layer]\;
    }
    }
    Performance, Energy, Area$\leftarrow$ Evaluator(optimal Mappings, inter-layer workloads)\;
    Evaluate design goal\;
    Update optimal Architecture\;
    }
\end{algorithm}
\label{alg:exploration}
\end{figure}

TRIM has an embedded energy and area model based on 65 nm technology. We utilize microarchitecture parameters such as bit width and PE pipeline stages; memory type, memory size, number of memory ports, wire length for NoC, etc., to compute each hardware component's energy and area. For the memory components, the model is based on CACTI\cite{muralimanohar2009cacti}. To improve the accuracy of our energy and area model, we also used the energy and area data from published papers\cite{venkataramani2017scaledeep, Chen2017b}. Both on-chip memory and off-chip memory are considered. The energy model can be replaced by other activity count-based energy models, such as Accelergy\cite{Wu2019}. The dynamic energy consumption is computed as the activities count multiply with the energy per activity. Both inter-layer and intra-layer workloads are considered. We also consider the network's sparsity, but we only consider zeros generated by padding and upsampling. The zeros come from input data, and ReLU cannot be predicted in hardware design. The static energy mainly comes from caching the intermediate activations. We get the time for caching each intermediate activation based on the estimated performance and compute the energy with the energy model's memory static power. The overall energy consumption is the combination of dynamic and static energy. The area is computed by adding the area of each hardware component listed in the hardware organization.

\subsection{TRIM Explorer}
The TRIM Explorer is the controller of the overall TRIM model. It utilizes different TRIM components to select the optimal architecture and corresponding optimal mappings based on the design goal, as shown in Algorithm 1. The task analyst generates the intra-layer workloads and inter-layer workloads. The designer creates the architecture space based on the hardware parameters. For each hardware description in the architecture space, we can find the global optimal mappings for each intra-layer workload with the exhaustive search method. This was used in all our case studies. Combining the optimal mappings and inter-layer workloads, we get the best performance and lowest energy consumption that the architecture can achieve.  Using this data, the explorer can fairly compare different architectures and select the optimal one.

\section{FPGA Validation And Case Study}\label{section:fpga}
\subsection{Experimental Setup}
To validate TRIM, we utilized it to explore an FPGA-based training architecture's design space. We selected AlexNet (AlexNet-Cifar)~\cite{Luo}, VGG-11 (VGG-Cifar)\cite{Simonyan2014a}, and ResNet-20 (ResNet-Cifar)\cite{He2016} and trained them with CIFAR-10 datasets\cite{Krizhevsky2009} to evaluate our system. The PYNQ-Z1 board was used as the FPGA platform. The board has a dual-core Cortex-A9 processor, a high-performance FPGA chip (Xilinx ZYNQ XC7Z020), and 512 MB DDR3 memory. To estimate time, energy, and area accurately, we adjusted model parameters based on the FPGA frequency, energy and resources utilization. Those data were collected by implementing and characterizing several essential components, such as MAC units and DMA channels on the FPGA.

\begin{figure}[!t]
\centering
\includegraphics[width=2in]{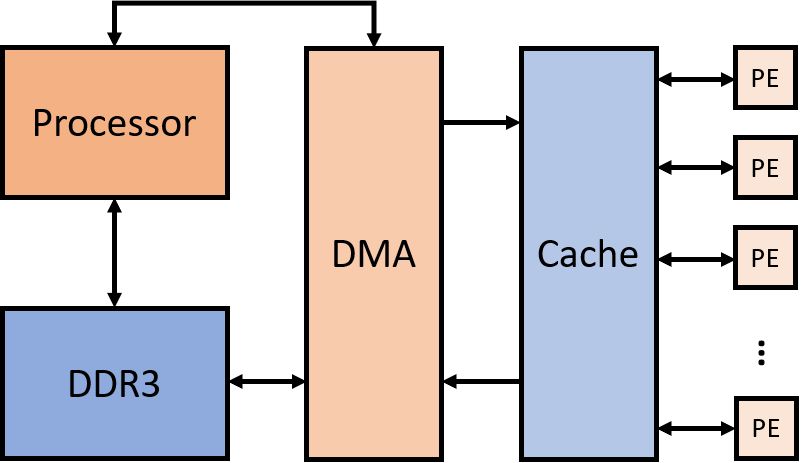}
\caption{Proposed FPGA Design}
\label{fig:fpga_design}
\end{figure}

Fig.~\ref{fig:fpga_design} shows the FPGA-based architecture we designed and implemented, which has both inference and training capabilities. The architecture has 32 parallel PEs and 32 KB cache. A DMA controller was used for high-performance data transfers between the accelerator cache and the DDR3 memory. The processor was used for dataflow control. The FPGA results achieved the same level of accuracy as the benchmarks trained on our PC with TensorFlow. To measure the real FPGA performance, we measured time using the Python time model. The energy was computed based on the FPGA power reported by Vivado and the time measured.

\subsection{Validate TRIM with Proposed FPGA Design}
\begin{figure}[!t]
\centering
\includegraphics[width=3in, height= 1.5in]{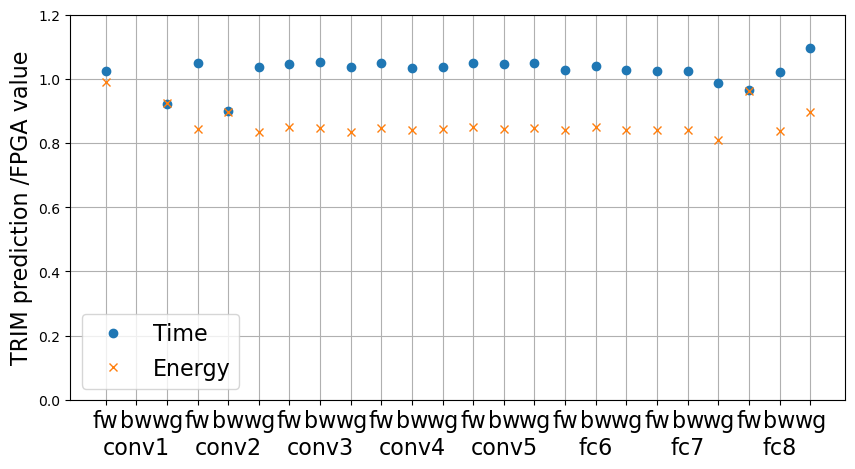}
\caption{Time accuracy for TRIM across all phase in AlexNet-Cifar}
\label{fig:t_validate_mAlex}
\end{figure}

\begin{figure}[!t]
\centering
\includegraphics[width=2.8in, height= 1.4in]{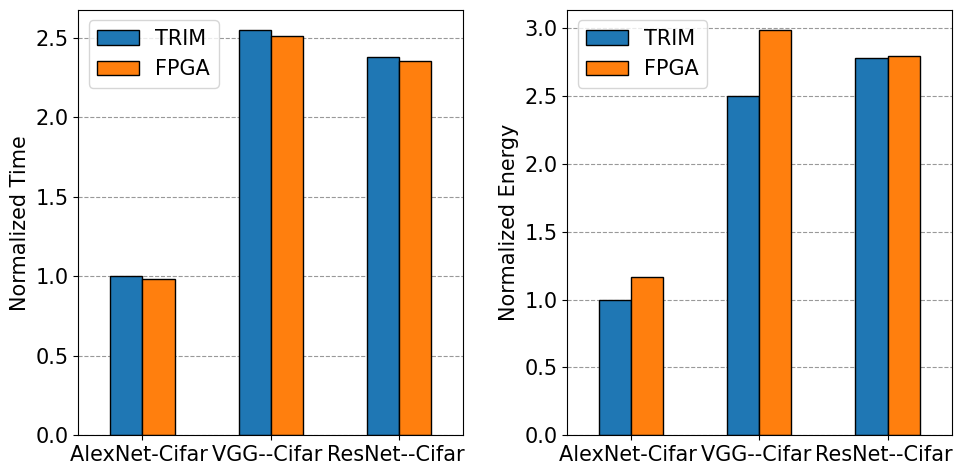}
\caption{Normalized time (left) and energy (right) comparison between TRIM and proposed FPGA}
\label{fig:te_validate}
\end{figure}

Fig.~\ref{fig:t_validate_mAlex} shows the TRIM prediction divided by the real value for the time and energy of each phase in AlexNet-Cifar on the FPGA. The overall time modeling error from TRIM was less than $10\%$ for each phase, while the energy modeling error was less than $20\%$. Fig.~\ref{fig:te_validate} shows the normalized time and energy of training our three benchmarks. The results show that TRIM accurately predicts training time and energy. As an early-stage model, TRIM offers good accuracy on both time and energy estimations.

\begin{table}[!t]
\renewcommand{\arraystretch}{1.3}
\caption{FPGA Exploration Hardware Setup}
\label{fpga_hw}
\centering
\begin{tabular}{|c|c|c|c|c|c|}
\hline
           & FPGA-1 & FPGA-2 & FPGA-3 & FPGA-4 & FPGA-5 \\ \hline
PE         & 8      & 16     & 32     & 64     & 128    \\ \hline
Cache (KB) & 20     & 24     & 32     & 48     & 80     \\ \hline
\end{tabular}
\end{table}

\begin{figure}[!t]
\centering
\includegraphics[width=2.8in, height= 1.4in]{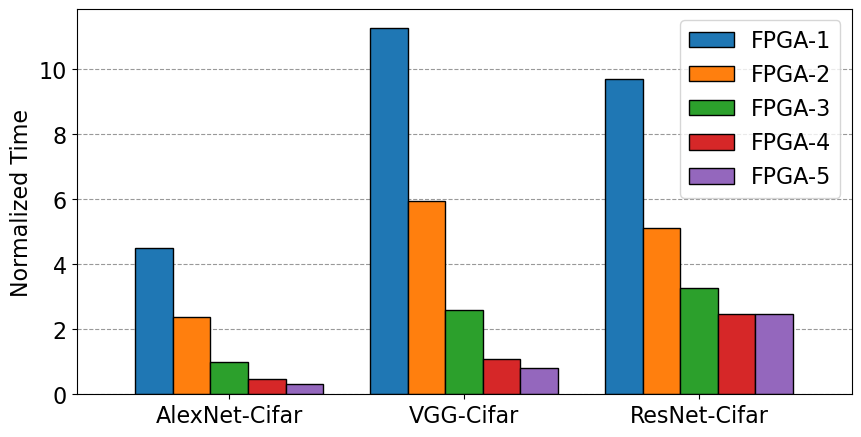}
\caption{Normalized time for selected hardware architectures using their own highest throughput mapping}
\label{fig:fpga_t}
\end{figure}

\begin{figure}[!t]
\centering
\includegraphics[width=2.8in, height= 1.4in]{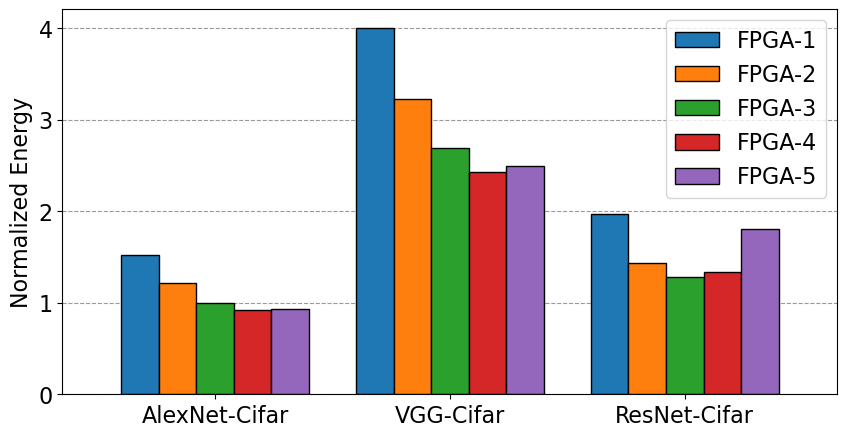}
\caption{Normalized energy for selected hardware architectures using their own highest throughput mapping}
\label{fig:fpga_e}
\end{figure}

\subsection{Explore Different FPGA Designs}
We varied the hardware parameters and configured five different FPGA hardware design points as shown in Table~\ref{fpga_hw}. For each design in Table~\ref{fpga_hw}, we searched for the mapping that provided the highest throughput. Fig.~\ref{fig:fpga_t} and Fig.~\ref{fig:fpga_e} show the normalized time and energy for our five different FPGA configurations with highest throughput for each of the three networks examined. As we increased the hardware resources used, the time needed to compute each network decreased. The only exception is with FPGA-5 training ResNet-20, where the number of PEs doubled compared to FPGA-4, but the throughput did not increase. This is because ResNet-20 has more layers, but each layer has far fewer parameters than the other two networks. This means that there is less scope for data parallelism.

For AlexNet, FPGA-5 and FPGA-4 consumed almost the same training energy, while FPGA-5 achieved a $1.38\times$ speedup. For VGG-11, FPGA-4 achieved the best energy efficiency, while FPGA-5 consumed $1.02\times$ more energy to achieve a $1.31\times$ speed up over FPGA-4. For ResNet-20, FPGA-3 achieved the best energy efficiency, while FPGA-4 consumed $1.04\times$ energy to achieve a $1.32\times$ speed up.

\begin{figure}[!t]
\centering
\includegraphics[width=2.5in]{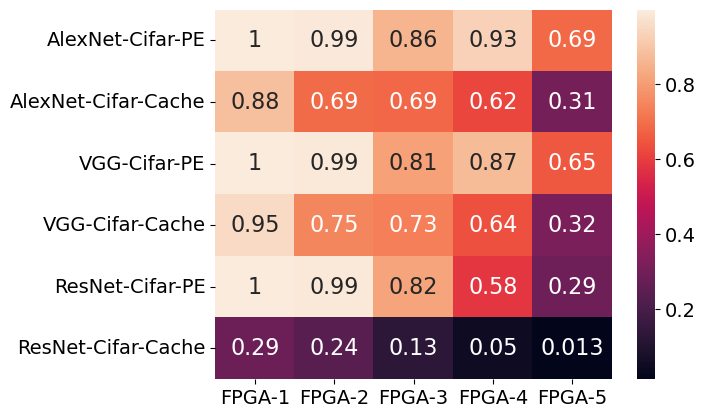}
\caption{FPGA hardware utilization rate}
\label{fig:fpga_util_rate}
\end{figure}

Fig.~\ref{fig:fpga_util_rate} shows the utilization rate of PEs and cache for each FPGA architecture training different networks. As FPGA-1 to FPGA-3 have a limited number of parallel PEs, the utilization rate of PEs is high for all three networks. FPGA-4 and FPGA-5 show higher PE utilization for AlexNet and VGG-11 than ResNet-20, which explains why FPGA-5 does not achieve higher training speedup for ResNet-20. The cache utilization for ResNet-20 is much less than the other two. The reason for that is ResNet-20 has much fewer parameters for each layer, which also requires much less memory space.

\begin{table}[!t]
\renewcommand{\arraystretch}{1.3}
\caption{FPGA Resources Results Generated by TRIM}
\label{fpga_resou}
\centering
\begin{tabular}{|c|c|c|c|c|c|}
\hline
     & FPGA-1 & FPGA-2 & FPGA-3 & FPGA-4 & FPGA-5   \\ \hline
LUT  & 9200   & 16400  & 30800  & 59600  & 117200   \\ \hline
FF   & 6300   & 11100  & 20700  & 39900  & 78300    \\ \hline
BRAM & 24     & 40     & 72     & 136    & 264      \\ \hline
DSP  & 40     & 80     & 160    & 320    & 640      \\ \hline
\end{tabular}
\end{table}

\begin{figure*} 
    \centering
  \subfloat[Time\label{fig:fpga_vt}]{%
       \includegraphics[width=0.3\linewidth]{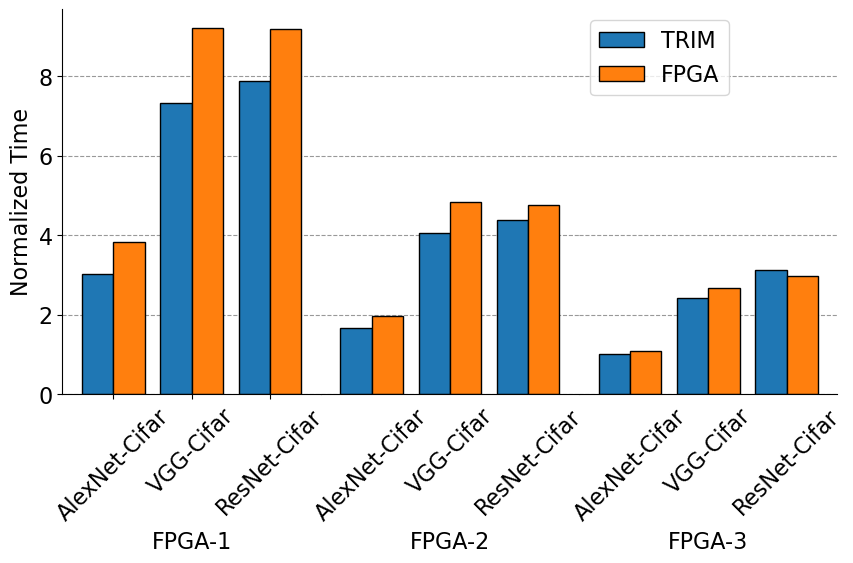}}
    \hfill
  \subfloat[Energy\label{fig:fpga_ve}]{%
        \includegraphics[width=0.3\linewidth]{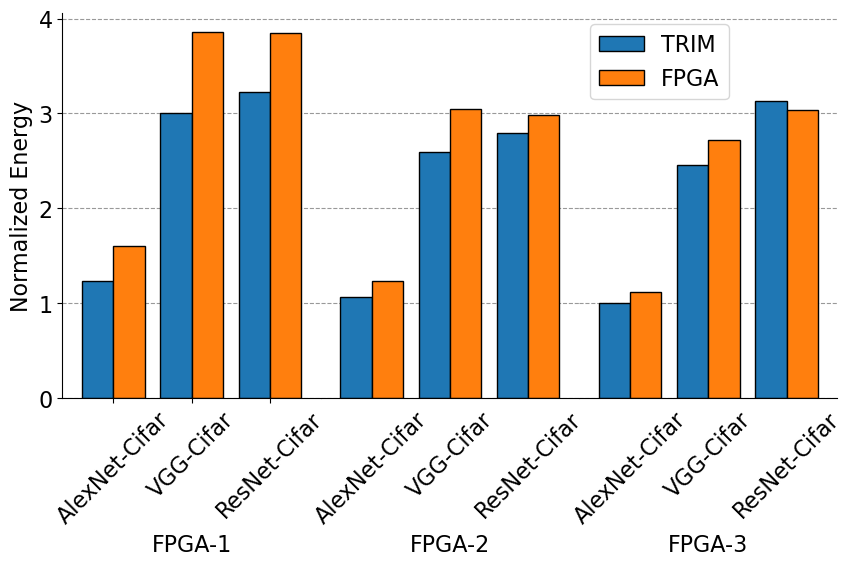}}
    \hfill
  \subfloat[Area\label{fig:fpga_va}]{%
        \includegraphics[width=0.3\linewidth]{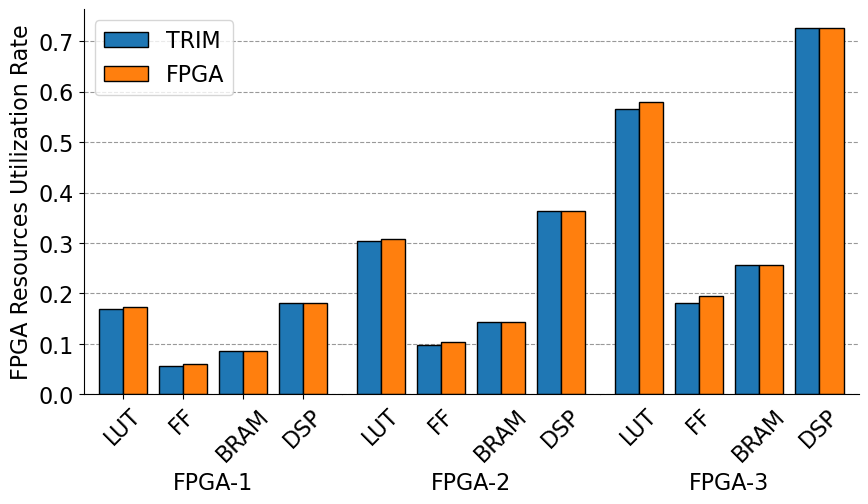}}
  \caption{Validate selected FPGA architecture}
  \label{fig:fpga_validate} 
\end{figure*}

\subsection{Validate TRIM with Different FPGA Designs}
Table~\ref{fpga_resou} shows the resources needed by the different FPGA designs predicted by TRIM. FPGA-4 and FPGA-5 required 320 and 640 DSP units respectively, while 220 DPS units were available on our PYNQ-Z1 board. Thus we could not implement FPGA-4 and FPGA-5 on the PYNQ-Z1 board. We implemented FPGA-1 to FPGA-3 on the PYNQ-Z1 board to validate with TRIM. 

Fig.~\ref{fig:fpga_vt} and Fig.~\ref{fig:fpga_ve} shows that TRIM accurately predicted the FPGA time and energy. As shown in Fig.~\ref{fig:fpga_va}, TRIM predicted LUT and FF usages within $5\%$ error while predicting BRAM and DSP usage accurately. Overall, TRIM accurately predicted the time, energy, and area.

\section{ASIC Validation and Case Study}\label{section:asic}
\subsection{Spatial architecture and TRIM validation}
\begin{figure}[!t]
\centering
\includegraphics[width=2.5in]{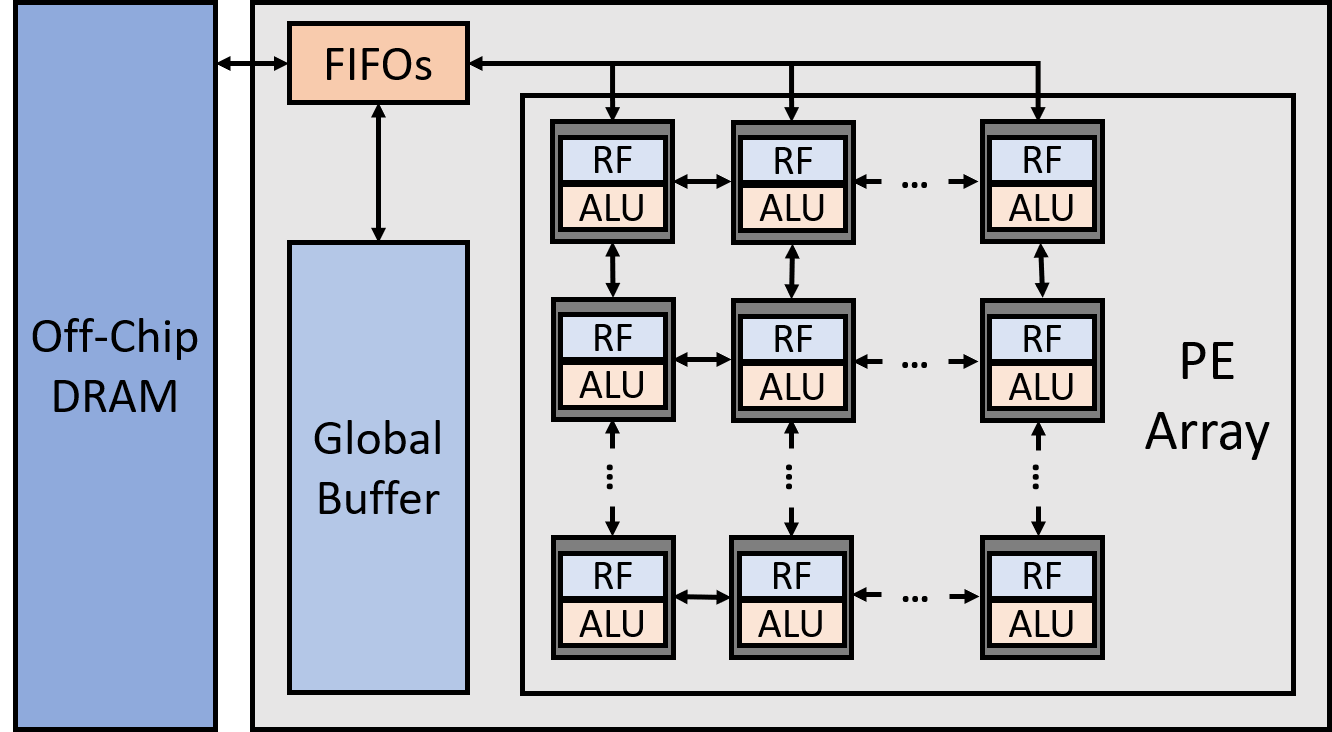}
\caption{Overview of proposed spatial architecture}
\label{fig:spatial_arch}
\end{figure}

This section shows how TRIM is used to explore the design space of spatial architectures. Fig.~\ref{fig:spatial_arch} shows an overview of the spatial architectures we are examining. These are widely utilized as DNN accelerators, especially for inference. They typically consist of tens to hundreds of simple PEs that communicate with each other through a NoC. These architectures show both throughput and energy efficiency on parallel computing tasks, such as convolution and matrix multiplications. Several groups\cite{Lu2017,Reagen2016} have designed unique spatial architectures to process the inference phase of DNNs efficiently.

Eyeriss~\cite{Chen2016} is a typical spatial architecture that is designed for the inference phase of DNNs. The original Eyeriss chip has a 108 KB on-chip global buffer (SRAM) and 168 PEs. Each PE has 512 bytes of registers. It utilizes a 16 bit data format, as this is enough for the inference phases of the deep networks considered.

\begin{figure} 
    \centering
  \subfloat[\label{fig:eyeriss_t}]{%
       \includegraphics[width=0.5\linewidth]{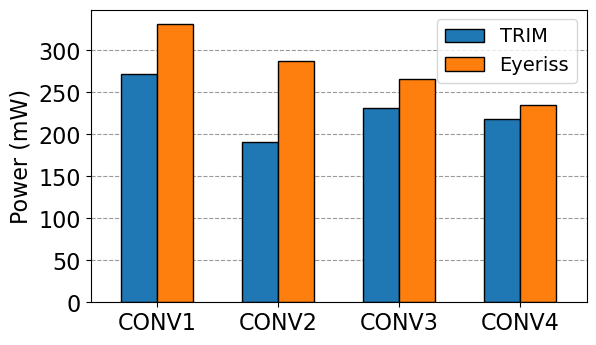}}
    \hfill
  \subfloat[\label{fig:eyeriss_e}]{%
        \includegraphics[width=0.5\linewidth]{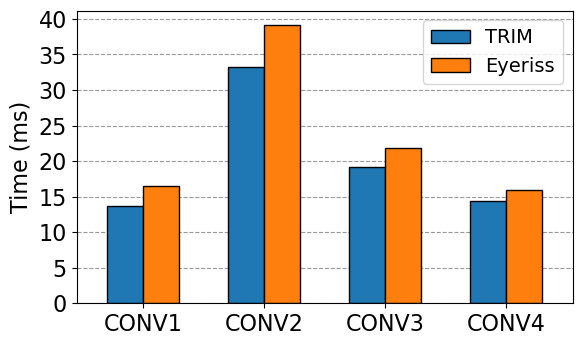}}
  \caption{Validate with Eyeriss inference accelerator}
  \label{fig:eyeriss_validate} 
\end{figure}

We validated our model with the Eyeriss, as properties for the hardware have been published for this accelerator. We selected the five convolutional layers listed in their paper for inference (Eyeriss does not do training). Fig.~\ref{fig:eyeriss_t} compares our TRIM predicted runtimes to the times listed in \cite{Chen2017b}. Although we overestimate the performance of Eyeriss, the most significant difference is for CONV1, with an error of about 17\%. Fig.~\ref{fig:eyeriss_e} shows that the power is underestimated by about 20\%, which was also the case with our FPGA power estimations.

We designed and modeled a spatial architecture with DNN training capability. We started with the Eyeriss’s hardware configuration and modified it to have training capability. We increased the data format to 32 bits as fewer bits will cause the training accuracy to drop. We added extra functional units to each PE to enable training: a transpose unit and a derivative unit based on the Eyeriss activation unit. We chose 256 PEs, 1024 Byte registers per PE, and a 256 kB Global buffer as our baseline hardware configuration.

Eyeriss uses a row stationary dataflow, which is one of the most energy efficient dataflows for inference processes. There are no published studies on which type of dataflow is best for training. In this study, we did not limit any mapping(dataflow) constraints and gave the model full capability to explore all possible dataflows. Thus TRIM can go through millions of possible valid mappings (dataflows) to get the best mapping for least energy consumption.

Five network models were selected as benchmarks: 1) AlexNet training ImageNet (AlexNet-IM)~\cite{c}, 2) AlexNet training Cifar-10 (AlexNet-Cifar)\cite{Luo}, 3) VGG-11 training ImageNet (VGG-IM)\cite{Simonyan2014a}, 4) ResNet-18 training ImageNet (ResNet-IM)\cite{He2016} and 5) MobileNet training ImageNet(MobileNet-IM)\cite{Howard2017}.

Below, we look at two case studies. In the first case, we use the baseline hardware configuration and study the effect on energy efficiency of sparsity circuits, different network models, datasets, and batch size. In the second case study, we use optimal options from case study I and vary the hardware configuration (size of register per PE and global buffer) to find optimal energy efficient hardware  configurations for 256, 512, and 1024 PEs architectures.

\subsection{Case Study I: Energy Efficiency Analysis for a Training Architecture}
\subsubsection{Utilize the sparsity of DNNs}
In DNN inference accelerator designs, data sparsity is utilized to achieve better energy efficiency.  The most commonly used method is to add circuits that skip operations with zero, including multiply and add by zero, which we named zero-skipping  circuits. For the inference phase of DNNs, the zeros come from two aspects: padding zeros in convolution computations and zero values in data. For the training of DNNs, there exists one more source: upsampling zeros, which are generated by upsampling operations shown in Eqs.~\ref{eq_wg} and ~\ref{eq_bw}. 

We estimated our baseline architecture's performance and energy with and without zero-skipping circuits. The zero-skipping circuits were implemented between the global buffer and register files. It is applied each time data is read from the global buffer to the register files. We only considered the padding zeros and upsampling zeros as they are determined by the network model and can be calculated without knowing the value of the input data. We want to note that as we didn't consider the zero values in data, the energy efficiency for the architecture with zero-skipping circuits is actually underestimated.

\begin{figure}[!t]
\centering
\includegraphics[width=2.8in, height= 1.4in]{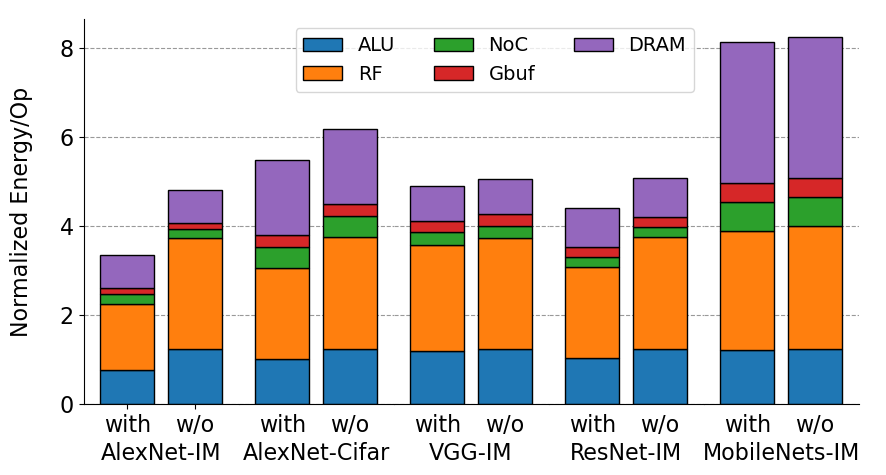}
\caption{Energy breakdown for different applications across each hardware level}
\label{fig:e_break_across_hw}
\end{figure}

\begin{figure}[!t]
\centering
\includegraphics[width=2.8in, height= 1.4in]{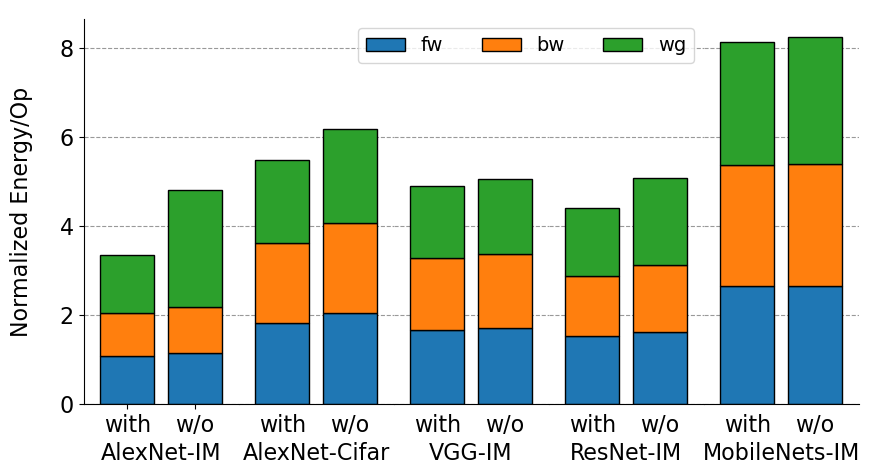}
\caption{Energy breakdown for different applications across each phase}
\label{fig:e_break_arcoss_phase}
\end{figure}

\begin{figure}[!t]
\centering
\includegraphics[width=2.8in, height= 1.4in]{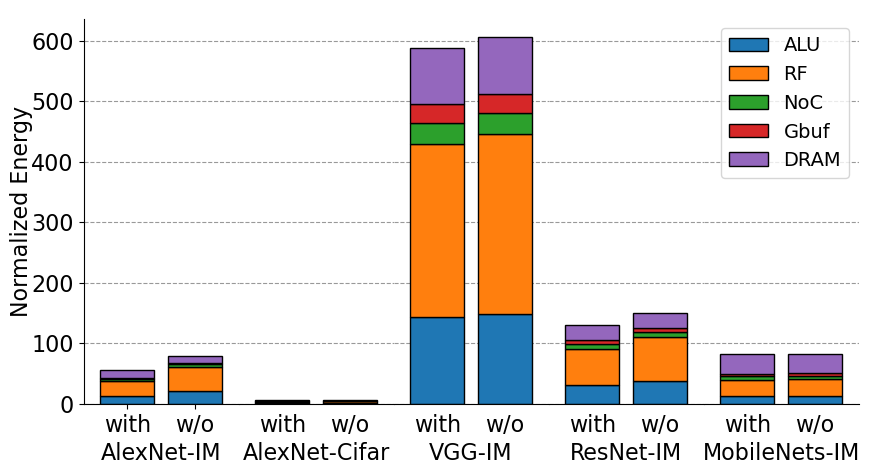}
\caption{Overall normalized energy for different networks}
\label{fig:overall_e_break}
\end{figure}

Figs.~\ref{fig:e_break_across_hw} and~\ref{fig:e_break_arcoss_phase} show the normalized energy per operation for our modeled architectures with and without skipping zero circuits for training. The architecture with zero-skipping circuits showed better energy efficiency for all four benchmarks, with the highest being 1.4x for AlexNet. As shown in Fig.~\ref{fig:e_break_across_hw}, the energy efficiency mainly comes from the ALU and register files. Fig.~\ref{fig:e_break_arcoss_phase} indicates that the energy efficiency primarily comes from the weight gradient phase, which means the energy efficiency mainly comes from the upsampling of propagation error. Thus the architecture with zero-skipping circuits achieves the best energy efficiency in training AlexNet, as it has the most upsampling operations among the four benchmarks. As the zero-skipping circuits can achieve energy efficiency with little area overhead, we implemented it in our following baseline architectures.

\subsubsection{Compare different applications}
Figs.~\ref{fig:e_break_across_hw} and~\ref{fig:e_break_arcoss_phase} show that the same architecture consumed different energy per operation while training various networks and datasets. Fig.~\ref{fig:e_break_across_hw} breaks down the energy per operation for different hardware components, while Fig.~\ref{fig:e_break_arcoss_phase} breaks down the energy per operation for different phases of training. Fig.~\ref{fig:overall_e_break} shows the normalized network energy for training one batch of 64 images broken down by hardware component. All the benchmarks are explored using our baseline hardware and best energy efficiency mapping criteria.

The different energy efficiencies come from different reuse opportunities of the input feature maps, filters, and output feature maps offered by the training datasets, network model, and different types of convolution operations. AlexNet-IM and AlexNet-Cifar utilized the same neural network architecture training different datasets. AlexNet-IM consumes more overall energy because of the larger image sizes, but consumes $1.7\times$ less energy per operation because the larger image sizes provide more filter reuse opportunities. In addition to datasets, different network architecture options lead to different energy efficiencies per operation. Among the four selected networks, AlexNet-IM training shows $1.41\times$ and $1.37\times$ energy efficiency per operation over VGG-IM and ResNet-IM respectively. This is because AlexNet utilized larger filter sizes, which could lead to more input feature maps and output feature maps reuse. MobileNet shows the most significant energy per operation as it utilizes depthwise convolution and pointwise convolution operations. Compared with the 2D convolution used in the other networks, these operations have far fewer data reuse opportunities and thus show the highest energy per operation.  However, these convolution methods significantly reduced the number of MAC operations, leading to higher overall energy efficiency than ResNet-IM and VGG-IM, as shown in Fig. ~\ref{fig:overall_e_break}.

\begin{figure}[!t]
\centering
\includegraphics[width=2.8in, height= 1.4in]{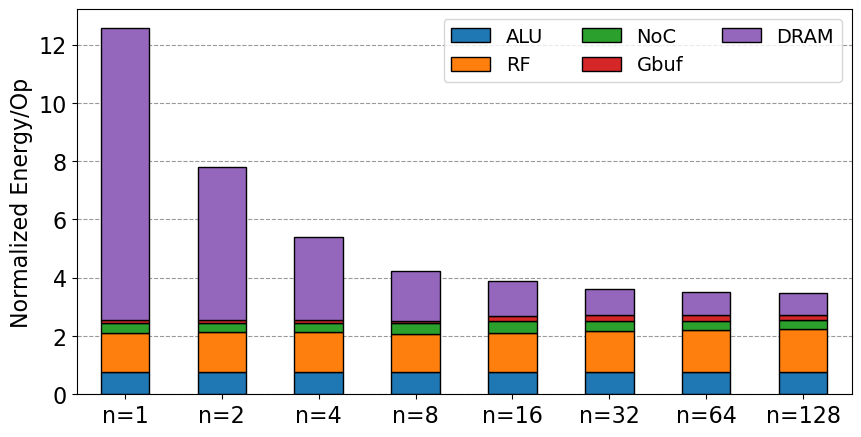}
\caption{Energy breakdown for different batch size training AlexNet-IM across each hardware}
\label{fig:e_batch_arcoss_hw}
\end{figure}

\begin{figure}[!t]
\centering
\includegraphics[width=2.8in, height= 1.4in]{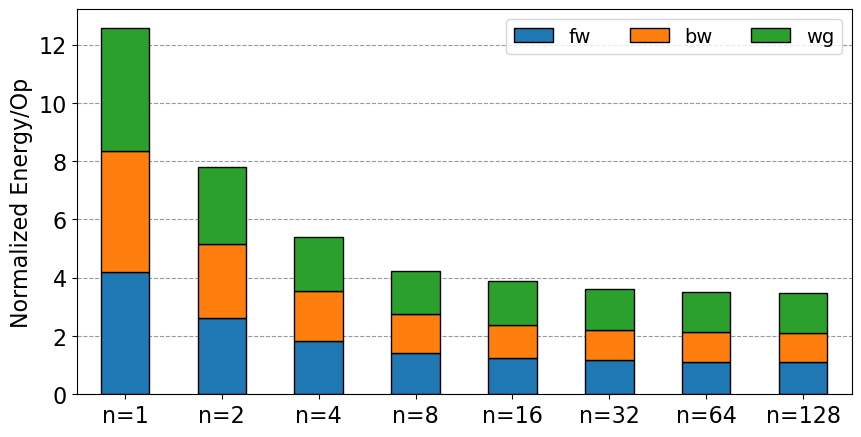}
\caption{Energy breakdown for different batch size training AlexNet-IM across each phase}
\label{fig:e_batch_arcoss_phase}
\end{figure}

\subsubsection{Batch size variation}
As shown in Figs.~\ref{fig:e_batch_arcoss_hw} and~\ref{fig:e_batch_arcoss_phase}, training with a larger batch size achieved better energy efficiency for the same architecture.  The differences come from the mappings utilized: the data with a larger batch size has more input feature maps and output feature maps reuse opportunities. TRIM can optimize the mapping to maximize the benefit of those reuse opportunities. Training with batch size 16 shows $3.1\times$ higher energy efficiency than training with batch size 1. The energy efficiency comes from reducing DRAM accesses and increasing accesses to global buffer and Network-on-Chip. However, training with batch size 128 shows negligible extra energy efficiency than training with batch size 64, even though it doubled the reuse opportunities of input and output feature maps. The bottleneck is the available hardware resources, which limits the reuse of each hardware level. Fig.~\ref{fig:e_batch_arcoss_phase} shows the energy efficiency evenly came from each phase, which means the batch processing can also be used for inference-only accelerators and achieve energy efficiency.

Overall, the network models, datasets, and batch size choices significantly affect the energy consumption for a given hardware architecture. In other words, if we want to compare the energy efficiency among different hardware architectures fairly, we must keep the network model, dataset, and batch size the same. 

\subsection{Case Study II: Optimize hardware architecture through design space exploration}
In this study, we build on the results from case study I and explore optimal architectures for the different networks. Based on the results of case study I, we implemented zero-skipping circuits and set the batch size to 64. We then varied the size of the register file per PE and global buffer to explore the design space for 256, 512, and 1024 PEs architectures. This was done for the AlexNet-IM, AlexNet-Cifar, and ResNet-IM benchmarks. The design goal was set to find the lowest energy-delay product options.

\begin{figure*} 
    \centering
  \subfloat[Normalized EDP of PE256 archi\label{fig25_i}]{%
      \includegraphics[width=0.3\linewidth]{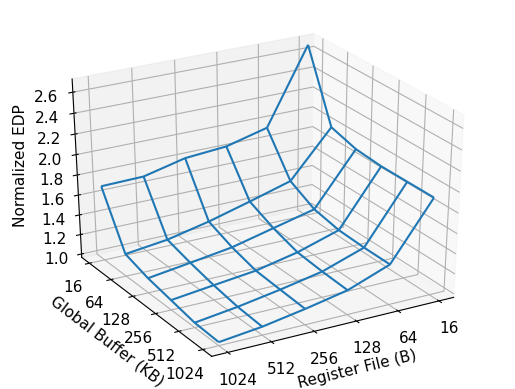}}
  \subfloat[Normalized EDP of PE512 archi\label{fig25_ii}]{%
      \includegraphics[width=0.3\linewidth]{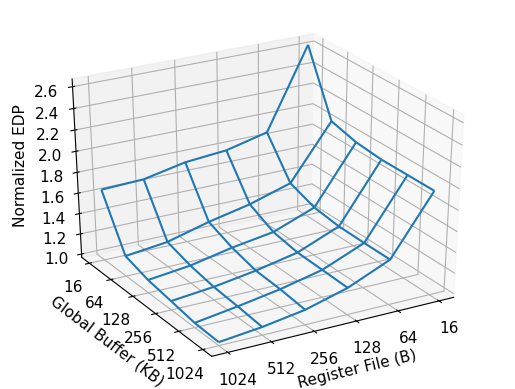}}
  \subfloat[Normalized EDP of PE1024 archi\label{fig25_iii}]{%
      \includegraphics[width=0.3\linewidth]{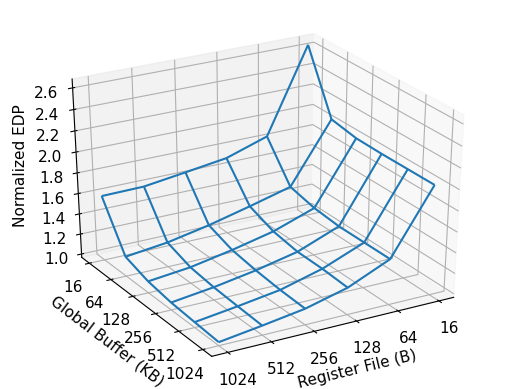}} \hfill
  \subfloat[Energy Efficiency of PE256 archi\label{fig25_iv}]{%
      \includegraphics[width=0.3\linewidth]{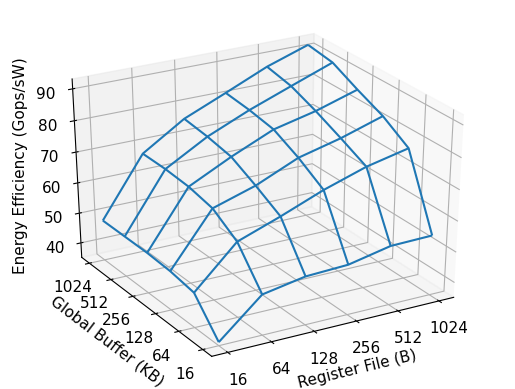}}
  \subfloat[Energy Efficiency of PE512 archi\label{fig25_v}]{%
      \includegraphics[width=0.3\linewidth]{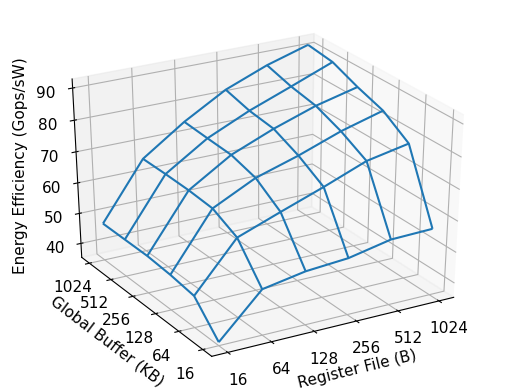}}
  \subfloat[Energy Efficiency of PE1024 archi\label{fig25_vi}]{%
      \includegraphics[width=0.3\linewidth]{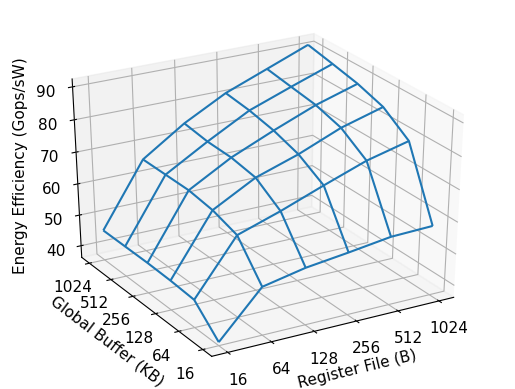}} \hfill
  \subfloat[Throughput of PE256 archi\label{fig25_vii}]{%
      \includegraphics[width=0.3\linewidth]{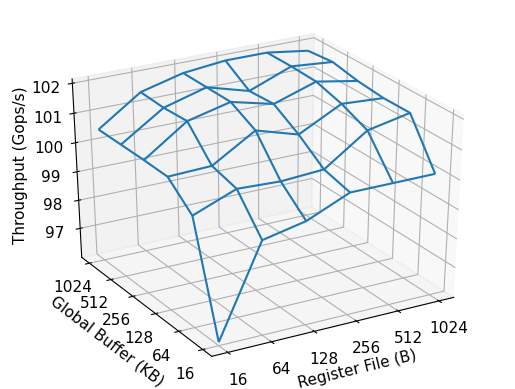}}
  \subfloat[Throughput of PE512 archi\label{fig25_viii}]{%
      \includegraphics[width=0.3\linewidth]{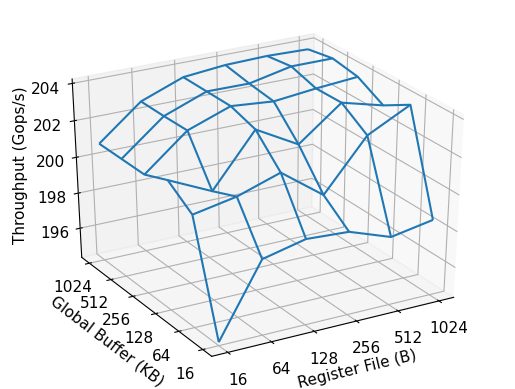}}
  \subfloat[Throughput of PE1024 archi\label{fig25_ix}]{%
      \includegraphics[width=0.3\linewidth]{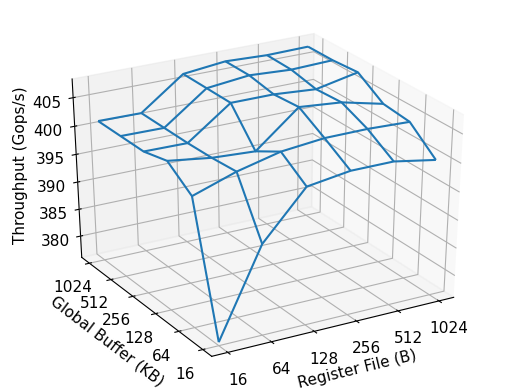}}
  \caption{Design space exploration results for AlexNet-Cifar with the design goal set as the lowest EDP.}
  \label{fig25} 
\end{figure*}

Fig.~\ref{fig25} shows the exploration results for AlexNet-Cifar with the design goal set as the lowest energy-delay product (EDP). As shown in Fig.~\ref{fig25_i} to Fig.~\ref{fig25_iii}, the EDP decreased with increases in hardware resources. Additional on-chip memory resources allow better caching of network parameters and intermediate activations, thus reducing the overall energy for off-chip memory accesses. We found that with more PEs, processing was faster, but the overall energy did not increase. Thus  additional PEs lowered the EDP. Thus a designer would have to trade off low EDP for additional chip area needed by PEs and memory. 

Fig.~\ref{fig25_iv} to Fig.~\ref{fig25_vi} show that for architectures with the same number of PEs, the larger the on-chip memory size (global buffer and register files), the lower the energy consumption. However, for architecture with different PE numbers but the same memory size, the energy consumption change is not that obvious. This indicates that the memory configuration is critical for the energy efficiency of the hardware. Fig.~\ref{fig25_vii} to Fig.~\ref{fig25_ix} show the throughput for different architectures. For architectures with the same number of PEs,  the processing time fluctuates with memory size increases. The reason for that is our design goal is the lowest EDP. TRIM traded off the processing time with energy to achieve the lowest EDP. Comparing the architectures with different numbers of PEs, we notice that the number of PEs is the key to achieving better performance. In the design space, the slowest architecture with 1024 PEs is 1.85x faster than the fastest architecture with 512 PEs. 

It is important to emphasize that an architecture's performance and energy efficiency does not rely only on the hardware resources available, but also on the mapping utilized. In our experiments, TRIM explored the mapping space and picked mapping that gave the lowest EDP for a given architecture. Simply scaling up the number of PEs without selecting a proper mapping would not significantly increase performance and energy efficiency.

\begin{figure}[!t]
\centering
\includegraphics[width=2.5in]{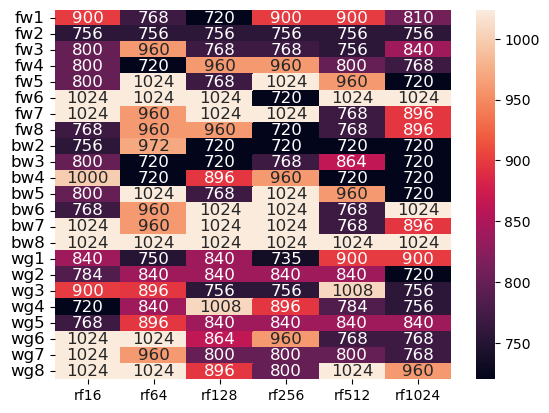}
\caption{Active PE heatmap 1024 PE architecture for AlexNet-Cifar (1024 KB global buffer)}
\label{fig:asic_heatmap}
\end{figure}

Fig.~\ref{fig:asic_heatmap} shows the number of active PEs for architecture with 1024 PEs,  1MB global buffer, and different register file sizes per PE. These results are for training AlexNet-Cifar. A unique number of PEs are activated for various architectures and network layers, which indicates that different mappings are used in specific architecture and network layers. Even when the same number of PEs are active, the mapping methods may be different in terms of memory usage. TRIM automatically explores the mapping space and finds the optimal approach to process the benchmark for different architectures based on the design goal.

Fig.~\ref{fig25} shows the design space exploration results for the AlexNet-Cifar benchmark. We have done a similar set of explorations for two other benchmarks: AlexNet-IM and ResNet-IM. In our design space explorations, we notice that we can achieve a smaller energy-delay product with more hardware resources. For instance, increasing the number of PEs produced significant speedups, while increasing the on-chip memory capacity allowed better energy efficiency. However, more hardware resources also mean a larger chip size. TRIM can quickly explore different hardware architectures and mappings to generate accurate performance and energy estimation.  This allows an architect to make hardware tradeoff decisions in the early stages of design.

\section{Conclusion}\label{section:conclusion}

We propose TRIM, an infrastructure for modeling and exploring the design space of DNN accelerators for both inference and training tasks. TRIM considers both intra-layer and inter-layer workloads to generate the estimation of performance, energy, and area. By considering inter-layer dependencies, TRIM can ensure the proposed hardware architectures have the capability to process all operations required for inference and training the DNN. In addition, the results are more reliable as they include the memory resources needed for caching the inter-layer data and the energy consumption of those inter-layer operations and data movements.We validate TRIM with the several FPGA designs and an ASIC-base inference accelerator (Eyeriss). After that, we demonstrate the usage of TRIM via case studies.

Future work includes extending TRIM to explore more state-of-the art deep networks, such as Transformer~\cite{Vaswani2017} and EfficientNet~\cite{Tan2019}. TRIM supports fixed precision quantization techniques. More quantization techniques such as mixed-precision quantization~\cite{Micikevicius2017} that require co-designed hardware will also be added in the future. In  recent  years,  neural  architecture search (NAS) utilizes machine learning models to design the network  architecture  and  to trade-off  between  accuracy  and efficiency. We would explore the possibility to combine TRIM with NAS to enable deep network and hardware co-design.

Overall, TRIM is a powerful tool for exploring the pros and cons in the hardware design space of DNN training accelerators for both FPGA and ASIC design. To the best of our knowledge, TRIM is also the first model which can explore the design space of training and inference DNN architectures.




 
\bibliography{trim_group}
%

\bibliographystyle{IEEEtran}

\begin{IEEEbiography}[{\includegraphics[width=1in,height=1.25in,clip,keepaspectratio]{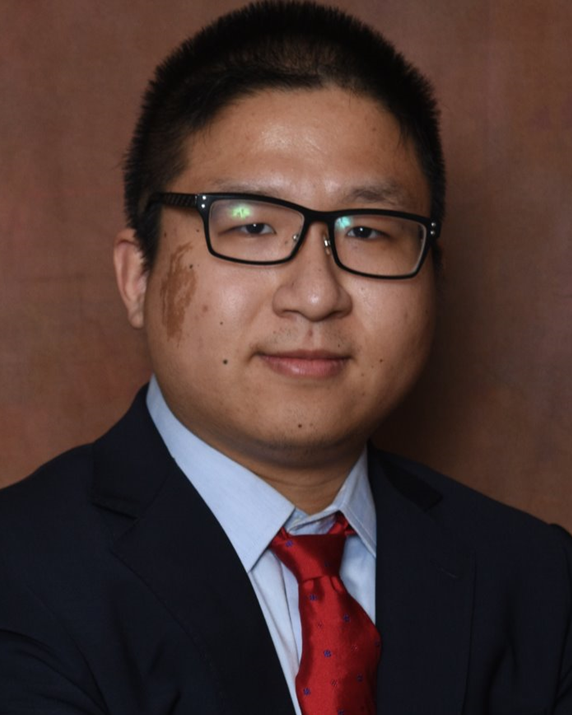}}]{Yangjie Qi}
(S'14) received the B.S. (2012) from the Department of Electrical Engineering, Anhui University, Hefei, China, and the M.S. (2015) from the Department of Electrical and Computer Engineering from the University of Dayton, OH. He is currently pursuing his Ph.D. at the University of Dayton. His Ph.D. research work focuses on low-power, high-performance architectures for deep learning. He is a student member of the IEEE.
\end{IEEEbiography}

\begin{IEEEbiography}[{\includegraphics[width=1in,height=1.25in,clip,keepaspectratio]{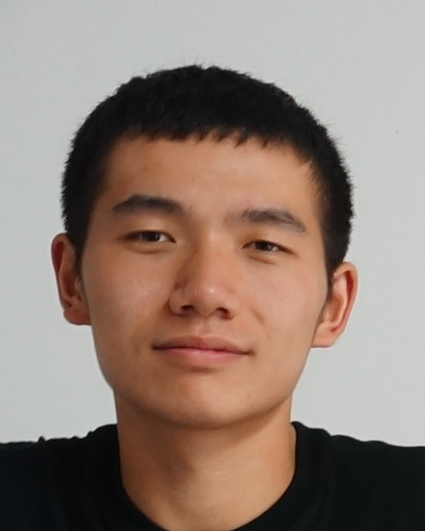}}]{Shuo Zhang}
(S'21) received the B.E. degree (2014) from Nanjing University of Science and Technology, Nanjing, China, and the M.S. degree (2016) from University of Dayton, Dayton, OH. He is currently working on his Ph.D in electrical engineering at University of Dayton. His research interests include the design and analysis of hardware architectures for deep learning and applications on the hardware. He is a student member of the IEEE.
\end{IEEEbiography}

\begin{IEEEbiography}[{\includegraphics[width=1in,height=1.25in,clip,keepaspectratio]{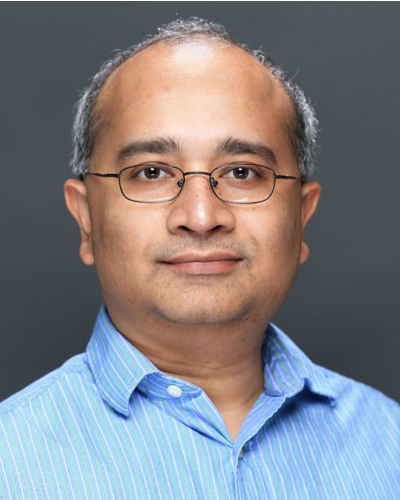}}]{Tarek M. Taha}
(S’96–M’03) is a Professor in the Electrical and Computer Engineering Department at the University of Dayton. He received the BS degree from DePauw University, Greencastle, Indiana, in 1996 and the BSEE, MSEE, and PhD degrees in electrical engineering from the Georgia Institute of Technology, Atlanta, in 1996, 1998, and 2002, respectively. His research interests include cognitive computing architectures, high performance computing, and architectural performance modeling. He received the NSF CAREER Award in 2007. His research is supported by multiple agencies and companies including the National Science Foundation, the Air Force Research Laboratory, and the National Aeronautics and Space Administration. He is a member of the IEEE and the IEEE Computer Society.
\end{IEEEbiography}



\vfill

\end{document}